\documentclass[10pt,doublecolumn, journal]{IEEEtran}
\IEEEoverridecommandlockouts

\usepackage{standalone}
\usepackage[reqno]{amsmath}
\usepackage{amssymb}

\usepackage{algorithm}
\usepackage{algorithmicx}
\usepackage{algpseudocode}
\usepackage{setspace}
\usepackage{multicol}
\usepackage{enumerate}
\usepackage[font=small, belowskip=-15pt,aboveskip=0pt]{caption}
\usepackage{cite}
\usepackage{amsthm}
\usepackage{float}

\usepackage[]{graphicx}
\usepackage{tikz}
\usepackage{tkz-tab}
\usepackage{pgfplots}
\usepackage{booktabs}
\usepackage{floatflt}

\usepackage{psfrag}	
\usepackage{array}
\usepackage{multirow,hhline}
\usepackage{exscale}
\usepackage{color}
\usepackage{colortbl}
\usepackage{epsfig}

\usepackage{array}
\usepackage[tight]{subfigure}

\usepackage[latin1]{inputenc} 
\usepackage[T1]{fontenc} 
\usepackage{rotating}
\usepackage{pbox}

\usepackage{pifont}

\usepackage{comment}
\usepackage{bookmark}
\usepackage{cleveref}
\usepackage{dblfloatfix}

\newcommand{\mat}[1]{\ensuremath{\boldsymbol{#1}}}
\renewcommand{\vec}[1]{\ensuremath{\boldsymbol{#1}}}

\DeclareMathOperator{\diag}{diag}

\algnewcommand\and{\textbf{and}}
\algnewcommand\Or{\textbf{ or }}

\algnewcommand\Switch{\State\textbf{switch }}
\newcommand{\Case}[1]{\State \textbf{case} #1\textbf{:} \begin{ALC@g}}

\newcommand{\EndCase}{\end{ALC@g}}
\algnewcommand\EndSwitch{\State\textbf{end switch}}

\usetikzlibrary{shapes,snakes}
\usetikzlibrary{calc}
\usetikzlibrary{patterns}
\usetikzlibrary{decorations.pathmorphing} 
\usetikzlibrary{spy}
\usetikzlibrary{positioning}
\usetikzlibrary{arrows, decorations.markings}
\usetikzlibrary{shapes.multipart, shapes.geometric, fit, backgrounds}
\usetikzlibrary{spy}

\definecolor{mycolor1}{RGB}{64, 122, 5}
\definecolor{mycolor2}{RGB}{97, 170, 209}
\definecolor{mycolor3}{RGB}{189, 0, 66}
\definecolor{mycolor5}{RGB}{230, 142, 0}
\definecolor{mycolor6}{RGB}{10, 10, 130}

\newcommand{\rev}[1]{\textcolor{black}{#1}}

\newcommand{\MU}[1]{{#1}}

\newcommand{\YK}[1]{{#1}}

\newcommand{\remove}[1]{}

\newcommand{\HC}[1]{\ensuremath{#1^h}}
\newcommand{\LC}[1]{\ensuremath{#1^l}}
\newcommand{\GC}[1]{\ensuremath{#1^c}}

\begin{document}

\title{Event-Based Framework for Agile Resilience in Criticality-Aware Wireless Networks}

\author{\IEEEauthorblockN{Yasemin Karacora\IEEEauthorrefmark{1}, Christina Chaccour\IEEEauthorrefmark{2}, Aydin Sezgin\IEEEauthorrefmark{1}, and Walid Saad\IEEEauthorrefmark{3}}\\
    \IEEEauthorblockA{\IEEEauthorrefmark{1} \small Ruhr University Bochum, Germany \\\IEEEauthorrefmark{2} Ericsson -Inc., Plano, Texas, USA\\\IEEEauthorrefmark{3} Bradley Department of Electrical and Computer Engineering, Virginia Tech, USA\\ Emails: \{yasemin.karacora, aydin.sezgin\}@rub.de, christina.chaccour@ericsson.com, walids@vt.edu \vspace{-.5cm}}
    \thanks{This work has received funding from the programme ``Netzwerke 2021'', an initiative of the Ministry of Culture and Science of the State of Northrhine-Westphalia, Germany. The sole responsibility for the content of this publication lies with the authors. W. Saad was supported by the Center for Assured and Resilient Navigation in Advanced TransportatION Systems (CARNATIONS) under the US Department of Transportation (USDOT)'s University Transportation Center (UTC) program (Grant No. 69A3552348324).}}

\maketitle

\begin{abstract}
As mission- and safety-critical wireless applications grow in complexity and diversity, next-generation wireless systems must meet increasingly stringent and multifaceted requirements. These systems demand resilience along with enhanced intelligence and adaptability to ensure reliable communication under diverse conditions.
This paper proposes an event-based multi-stage resilience framework for systematically integrating complementary error mitigation techniques in wireless networks.
The framework is applied to uplink transmission of mixed-criticality data under random link blockages. 
A key component is a novel mixed-criticality rate-splitting multiple access (MC-RSMA) scheme that combines multi- and single-connectivity to balance rate and blockage robustness. MC-RSMA is complemented by one-sided access point cooperation and central decoding, which are integrated into an event-driven algorithm. Here, increasingly effective but more complex mechanisms are activated sequentially to systematically counteract blockages while balancing resilience with cost.
From a cross-layer perspective, two transmit power allocation problems are formulated: One for separate decoding and one for central decoding, to ensure fair queue utilization under heterogeneous quality-of-service requirements.
Extensive simulations are used to evaluate the delay performance under varying blockage durations and examine the cost tradeoffs among resilience mechanisms within the proposed framework.
Results show that the proposed framework achieves resilience across disruption regimes: MC-RSMA balances efficiency and robustness as a criticality-aware core scheme, active robustness strategies handle frequent short-term fluctuations, and adaptive recovery ensures performance during rare, prolonged blockages.
\end{abstract}


\linepenalty=1000
\vspace{-2pt}
\section{Introduction}
Next-generation wireless systems must support a plethora of new applications characterized by diverse quality-of-service (QoS) requirements. These demands encompass not only high data rates, but also extreme reliability and minimal latencies, which are essential in applications such as extended reality (XR), vehicular communications, and networked control. These services are envisioned for use cases with high criticality levels, spanning industries like healthcare, intelligent transportation, and industrial manufacturing \cite{saad20206G}. In order to effectively serve such use cases, 6G communication must reliably ensure strict adherence to the prescribed QoS requirements, since any deviation not only impacts service functionality, but also carries the risk of serious consequences, including threats to human safety and potential environmental harm.\looseness-1

Resilience has always been a cornerstone of communication networks for maintaining security and service continuity. However, new challenges emerge as mission-critical use cases grow in complexity and intelligence, requiring networks to navigate conflicting demands within increasingly cloud-based, virtualized, and autonomous architectures. 
While 5G focused on reliability (preventing failures) and robustness (maintaining stability during disturbances), 6G research elevates \emph{resilience} as a core design principle \cite{nextGAroadmap, IMT-2030}. Resilience in 6G extends beyond withstanding disruptions; it emphasizes rapid adaptation and recovery to maintain high-performance connectivity despite inevitable failures \cite{madni2009resilience}. Emerging 6G technologies, such as higher frequency bands and massive MIMO, amplify the need for resilience by design to address new vulnerabilities and demands in mission-critical services.

These advanced physical layer techniques, designed to meet high data rate demands, use narrower beams, which increases sensitivity to line-of-sight (LoS) blockages and deep fading, particularly in dynamic environments \cite{maccartney2017blockage}.
Thus, ensuring stable service functionality despite channel fluctuations requires carefully designed countermeasures.
Here, existing solutions show a fundamental tradeoff: permanent redundancy and diversity, e.g., multi-connectivity and cooperation, improve reliability but waste resources, while frequent adaptations improve efficiency but introduce overhead and delay, which is unsuitable for mission-critical scenarios. A key challenge is therefore to balance service availability and resource efficiency through more agile resilience strategies \cite{khaloopour2024resilience, mahmood2024resilient}.

Consequently, resilient and time-critical communication schemes should be designed to incorporate diverse robustness and recovery mechanisms. These include a variety of proactive and reactive strategies, that are deployed in a targeted and balanced fashion according to the likelihood and potential risks of errors. 
For instance, disturbances anticipated to occur with high probability, like temporary deep fades and channel uncertainties, may require mitigation strategies based on inherent redundancy and diversity, e.g., multi-antenna configurations \cite{perez2023multiAntenna} or frequency diversity \cite{aboagye2024multiband}. This approach helps avoid frequent communication disruptions and the computational effort and delays caused by constant adjustments to the transmission scheme. Conversely, rare or less harmful errors are better addressed with adaptation and recovery mechanisms, since the permanent occupation of resources would be inefficient in such cases.

To ensure that appropriate countermeasures are taken whenever needed, \emph{event-based} algorithms are a promising approach for efficient and flexible adaptation to heterogeneous conditions. For instance, 3GPP already introduces event-driven and user-initiated procedures in beam management for 5G-Advanced in Release 19 \cite{lin2023Rel19}. 
Such event-driven adaptation allows networks to activate countermeasures only when necessary, ensuring both agility and efficiency in resilience management.

Beyond adapting to channel events, resilience also requires differentiation across services with diverse QoS requirements and criticality levels.
 With increased connectivity, diverse applications compete for limited network resources to optimize overall user experience. Unexpected channel fluctuations and blockages can disrupt service, challenging continuous operation guarantees. In such scenarios, service prioritization becomes indispensable. 
 A common approach is network slicing, where each service is assigned dedicated resources. However, static isolation in orthogonal slicing restricts dynamic resource sharing, hindering flexible adaptation to changing network conditions. Therefore, achieving resource-efficient resilience calls for more agile allocation strategies, such as non-orthogonal schemes, which enable dynamic sharing of resources without sacrificing service guarantees.

\subsection{Prior Art}
The concept of \emph{resilience}, traditionally explored in psychology, ecology, or social sciences, has recently gained significant attention in wireless communication research, particularly in 6G development \cite{nextGAroadmap, IMT-2030}. 
Although communication schemes addressing robustness are well-established, there is a growing need for integrated approaches that combine proactive robustness with reactive response mechanisms.
Existing resilience frameworks \cite{madni2009resilience, khoury2014multi, punzo2020resilience}, and their application to communication networks \cite{sterbenz2010resilience, kaada2022resilience, reifert2022comeback, shui2024design, mahmood2024resilient, khaloopour2024resilience}, cover different resilience phases, such as error detection, defense, remediation, and recovery. 
Notably, the works in \cite{mahmood2024resilient} and \cite{khaloopour2024resilience} emphasize the importance of \emph{efficient} resilience strategies that balance cost-performance tradeoffs and incorporate criticality-awareness.
These works underline the need for tailored approaches that address the unique demands of diverse use cases, rather than relying on a universal solution. 
In this context, the authors in \cite{reifert2022comeback} explore criticality-aware resource management by comparing multiple resilience mechanisms with varying effectiveness and complexity. The work in \cite{li2023trafficresilience} emphasizes the need for resilience strategies to consider both the frequency and impact of anomalies.
However, these studies lack focus on the design of control algorithms for activating appropriate failure responses.

Robustness and resilience in the context of \emph{channel blockage} and LoS intermittence have been widely studied \cite{gerasimenko2019multiconn, rahman2019beamswitch, kumar2021CoMP, reifert2023CoNOMA, barbarossa2019resilient, karacora2024intermittency, karacora2024THzRIS}. 
An effective approach to overcome link blockages is through the use of spatial macro-diversity along with multi-connectivity and cooperative transmission, as explored in various works, including \cite{gerasimenko2019multiconn, rahman2019beamswitch, kumar2021CoMP, reifert2023CoNOMA, barbarossa2019resilient}.
For example, the authors in \cite{gerasimenko2019multiconn} address the complexity of multi-connectivity, while the work in \cite{rahman2019beamswitch} proposes beam-switching to handle time-varying blockages in mmWave channels.
The work \cite{kumar2021CoMP} utilizes coordinated multi-point (CoMP) transmission and robust beamforming to handle blockages. In \cite{reifert2023CoNOMA}, the authors introduce a cooperative non-orthogonal multiple access (NOMA) approach to serve user devices with varying channel conditions, effectively countering blockages in extended reality applications.\looseness-1

Most works on multi-connectivity and cooperative communication schemes, including \cite{gerasimenko2019multiconn, rahman2019beamswitch, kumar2021CoMP, reifert2023CoNOMA}, focus on downlink transmission. However, emerging 6G use cases like XR applications, mobile edge computing (MEC) offloading, and massive IoT data gathering, will require more intense uplink traffic, demanding high throughput, reliability and low latencies \cite{bassoli2021why6G}.
Blockage resilience in the uplink has been addressed in several works, e.g., via multi-link computation offloading for MEC applications \cite{barbarossa2019resilient}, or CoMP reception schemes \cite{maham2020CoMPuplink}.
The authors in \cite{abbasi2022transmission} investigate user cooperation, proposing an uplink rate-splitting multiple access (RSMA) scheme, in which users exchange their transmit signals and forward them to the base station in sequential time slots.
Despite numerous works proposing strategies for handling blockages (such as RIS, cooperation, and non-orthogonal access schemes), these studies often focus on individual techniques in isolation. There is, however, a notable gap in research regarding the effective integration of diverse countermeasures against blockages within a cohesive resilience framework.

One crucial factor to be incorporated into resilient communication schemes is \emph{criticality-awareness}, as it can substantially enhance energy efficiency (e.g., \cite{reifert2022MC_RSMA}), and compliance with strict requirements for safety-critical applications \cite{reifert2023AoI, park2021nonlinearAoI}. 
While orthogonal network slicing is a common but resource-intensive approach to serving mixed-criticality applications simultaneously, \cite{popovski2018H_NOMA} proposes non-orthogonal slicing, known as heterogeneous NOMA (H-NOMA), which has been shown to be superior in certain regimes. H-NOMA leverages diverse reliability requirements by employing successive interference cancellation (SIC) and decoding critical data first.
Similarly, in \cite{karacora2024THzRIS}, a superposition coding scheme for RIS-aided THz systems is introduced, which ensures high reliability for critical data, while efficiently transmitting non-critical data when a LoS link is available.
As a generalization of NOMA, RSMA \cite{mao2022rate} has gained interest for its efficiency and resilience benefits in both downlink and uplink scenarios \cite{abbasi2022transmission, yang2019uplinkRSMA, reifert2022comeback, ahmad2019BSbreakdown, karacora2022ratesplitting}. 
Unlike H-NOMA, RSMA involves message splitting and thereby can enable users to simultaneously transmit multiple data streams with heterogeneous reliability levels. As in RSMA a common message stream is decoded at multiple receivers, it can be used to support multi-connectivity for highly critical data, while transmitting less critical data on private streams, thereby enhancing reliability in a resource-efficient manner. In our previous work \cite{karacora2022ratesplitting}, we proposed a novel RSMA-based scheme designed to enable hybrid multi- and single-connectivity for criticality-aware uplink transmission with finite blocklength coding.
Here, we build upon this RSMA framework, generalizing it to multi-user scenarios and embedding it into a three-stage resilience strategy to effectively manage LoS blockages.

In summary, while foundational resilience frameworks and criticality-aware strategies exist, there remains a significant gap in a holistic approach that balances robustness, efficiency, and adaptability in mixed-criticality uplink communication.

\subsection{Contribution}
\begin{figure}
    \centering
    \includegraphics[width=0.9\linewidth]{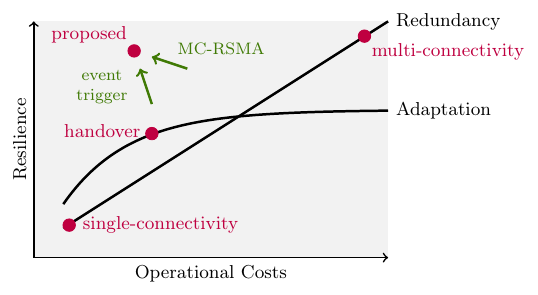}
    \caption{Conceptual illustration of the contribution compared to common approaches. Single-connectivity offers low resilience, while permanently redundant schemes such as full multi-connectivity incur high operational costs. Adaptation-based mechanisms, e.g., handovers, are limited by delay. The proposed event-driven, criticality-aware hybrid scheme achieves a more efficient balance between resilience and cost.}
    \label{fig:contribution}
\end{figure}
The main contribution of this paper is a novel resilience framework that effectively combines robustness- and adaptation-oriented techniques in the context of wireless communication networks\footnote{We consider resilience at the network and system level, specifically in adapting to and recovering from disruptions. Aspects related to security, such as defense against jamming, are beyond the scope of this paper.}. As illustrated in Fig. \ref{fig:contribution}, redundancy-based methods are cost-intensive, whereas adaptation alone is limited by delays and overhead. The proposed framework addresses this fundamental tradeoff by integrating criticality awareness, non-orthogonal resource sharing, and event-driven scheme switching to achieve high resilience in a cost-efficient manner.
We specifically consider uplink transmission\footnote{This work focuses on uplink transmission as meeting the uplink traffic demand is predicted to be a bottleneck for many 6G applications \cite{bassoli2021why6G}, such as XR or vehicular communications. Note that while downlink is outside the scope of this work, similar resilience concepts can be applied in the downlink, e.g., using coordinated multipoint transmission and soft handover frameworks.} affected by LoS blockage and propose an event-driven resilience algorithm involving a novel RSMA-based multi-connectivity scheme and access point (AP) cooperation to counteract link intermittency. 
Our key contributions include:
\begin{itemize}
    \item We introduce a \emph{three-stage resilience framework} for uplink LoS blockage scenarios, where more effective but complex countermeasures are gradually activated when needed. In contrast to prior work focusing on isolated solutions, our event-driven and criticality-aware design provides a novel, integrated approach that balances resilience, resource consumption, and coordination costs (see Fig. \ref{fig:contribution}).
    \item Adopting a cross-layer perspective, we propose a \emph{criticality-aware transmission scheme}, where each user's data is modeled via two queues with distinct criticality levels. In Stage 1 of the proposed resilience framework, being deployed in normal operation mode, we propose a novel RSMA-based uplink scheme that leverages message splitting to enable a hybrid multi-/single-connectivity approach: critical data benefits from multi-connectivity to ensure blockage robustness, while low-priority data is transmitted via single-connectivity to maintain throughput efficiency.
    \item Within our event-driven resilience strategy, we propose enabling AP cooperation as needed to maintain service functionality despite blockages. In Stage 2 of the multi-stage framework, we consider one-sided AP cooperation as an \emph{active} robustness strategy, that can partially absorb the performance degradation caused by LoS blockage.
    We further consider full AP cooperation with central decoding in Stage 3, where the APs forward their received signals to the central unit (CU) for joint processing.
    \item We formulate power allocation optimization problems for both separate and central decoding, that aim at stabilizing the system while prioritizing critical traffic during blockages. The non-convex problems are solved iteratively using successive convex approximation and fractional programming techniques.
    \item We provide extensive simulations under varying blockage statistics, offering insights into the effectiveness and role of the different resilience mechanisms. Results demonstrate that the proposed MC-RSMA outperforms orthogonal, single-, and full multi-connectivity baselines by sustaining minimal delays for critical traffic. While AP cooperation stabilizes low-criticality traffic, power optimization becomes essential for longer blockage durations, which highlights the dependence of effective countermeasures on error dynamics. The event-driven algorithm is shown to balance resilience and cost more effectively than single-strategy designs.
\end{itemize}
The rest of this paper is structured as follows. Section \ref{sec:sys_model} introduces the system model for a multi-user uplink transmission scenario with LoS blockages. Section \ref{sec:resilience} proposes a three-stage resilience framework. Section \ref{sec:MC_RSMA} introduces the criticality-aware multi-connectivity scheme based on RSMA, while Section \ref{sec:coop} details the cooperation schemes. Section \ref{sec:event_alg} describes the event-driven resilience algorithm. Numerical results are presented in Section \ref{sec:results} followed by conclusions in Section \ref{sec:conclusion}.

\section{System Model}\label{sec:sys_model}
\begin{figure}
    \centering
    \includegraphics[width=0.94\linewidth]{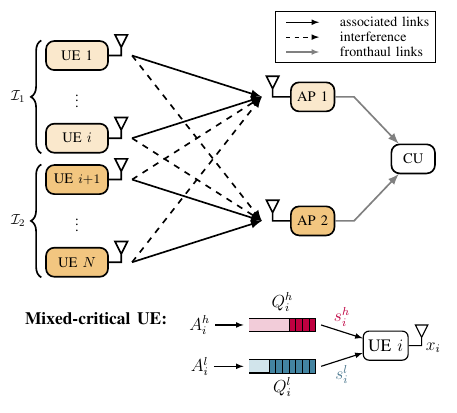}
    \caption{\YK{System model comprising $N$ UEs with mixed-criticality data queues and two APs.}}
    \label{fig:sys_model}
\end{figure}

We consider the uplink communication of a set $\mathcal{I}$ of $N$ single-antenna user equipment (UE) connected to two APs, which are both linked to a CU as illustrated in Fig. \ref{fig:sys_model}. For simplicity, we focus on a scenario with two APs.\footnote{As we consider non-orthogonal RSMA-based transmission, note that by applying user grouping, interference from users associated with other APs can be managed via orthogonal multiple access methods \cite{mao2022rate}. Therefore, we limit our study to the two AP case for tractability.} 
Each user generates mixed-criticality data for uplink transmission, where data packets are categorized in two different criticality levels, i.e., \emph{high criticality} (HC) data with strict latency and reliability requirements, and \emph{low criticality} (LC) data, where delayed delivery or discarded packets do not have such severe consequences. Mixed-criticality streams may involve different data types, such as critical sensor data for control or vehicular applications versus less critical monitoring data, or a single data type divided into layers, as in scalable video or region-of-interest coding.
This separation of data packets into streams of different criticality allows for a tailored transmission strategy that can ensure uninterrupted connectivity for HC data during disturbances and thereby enables dependable operation of mission-critical applications.

\subsection{Channel Model}
We assume that the channel between UE $i$ and AP $j$ is subject to Rician fading with a dominant LoS path, which is intermittently blocked, and a non-line-of-sight (NLoS) component with continuous availability. We consider dynamic LoS blockage caused by moving objects and obstacles that is modeled as a Markov process with Poisson blockage arrivals and exponential blockage time. 
The channel gain of the link between UE $i$ and AP $j$ is given as 
\begin{equation} \label{channel}
    h_{ji} = \sqrt{\mathrm{PL}_{ji}} \left(\beta_{ji} \sqrt{\frac{K}{K+1}} + \sqrt{\frac{1}{K+1}} h_\mathrm{NLoS}\right),
\end{equation}
where $\mathrm{PL}_{ji}$ is the path loss of the channel between UE $i$ and AP $j$, $K$ represents the Rician $k$-factor, i.e., the power ratio of the LoS and NLoS channel component, and the NLoS channel is modeled as $h_\mathrm{NLoS} \in \mathcal{CN}(0,1)$. The LoS component is determined by the binary variable $\beta_{ji} \in \{0,1\}$, modeled as an alternating renewal process, such that the blockages occur following a Poisson process with arrival rate $\kappa_\mathrm{B}$ blockers/sec and the blockage duration is assumed to be exponentially distributed with parameter $\mu_\mathrm{B}$ 1/s \cite{jain2019block}.
Hence, the blocking and unblocking probabilities are given as\looseness-1
\begin{equation} \label{blockage_prob}
\begin{split}
    &P\left(\beta_{ji}(t)=0 | \beta_{ji}(t-1)=1\right) = 1 - e^{-\kappa_\mathrm{B} T},~ \text{and} \\&P\left(\beta_{ji}(t)=1|\beta_{ji}(t-1)=0\right) = 1 - e^{-\mu_\mathrm{B} T}.
    \end{split}
\end{equation}
The overall probability that the LoS link between UE $i$ and AP $j$ is blocked in time slot $t$ will be: 
\begin{equation}
   p_\mathrm{b} = P\left(\beta_{ji}(t) = 0\right) =  \frac{\kappa_\mathrm{B}}{\kappa_\mathrm{B} + \mu_\mathrm{B}}.
\end{equation}

We assume that channel state information (CSI) measurements are performed at the beginning of each time slot, so that $h_\mathrm{NLoS}$ is perfectly known by the users. However, a disruption of the LoS path can occur during a time slot after the CSI measurement has already been obtained. Hence, we assume that LoS blockage is detected in the time slot following its initial occurrence (with a delay of one time slot).
Defining $\hat{\beta}_{ji}$ as the expected blockage state, we have
\begin{equation}
    \hat{\beta}_{ji}(t) = \begin{cases}
        1, & \text{if } \beta_{ji}(t-1)=1 \text{ and } \beta_{ji}(t)=0,\\
        \beta_{ji}(t), & \text{otherwise}.
    \end{cases}
\end{equation}
Note that for the scope of this work, we consider LoS blockage as the only source of channel uncertainty. Thus, the expected channel state $\hat{h}_{ji}$ at time slot $t$ can be obtained from \eqref{channel} with $\beta_{ji}=\hat{\beta}_{ji}$. Furthermore, we neglect user mobility by accounting only for small-scale Rician fading, while assuming constant path loss. We also assume that all considered UEs in $\mathcal{I}$ remain within communication range of both APs. Note that with UE mobility, handovers and user grouping adjustments would become necessary, which are not considered in this work for the sake of tractability. We further assume that the UE-AP association is determined based on the LoS path loss, with $\mathcal{I}_j \subset \mathcal{I}$ representing the set of user indices associated with AP $j$, so that $\mathcal{I}_j = \{i\in\mathcal{I} \mid \mathrm{PL}_{ji} > \mathrm{PL}_{j'i}, j'\neq j\}$.

\subsection{Queuing Model}
To model the mixed-critical data at the users, we adopt a M/G/1 queuing system\footnote{As we apply an RSMA scheme, where the HC and LC data streams are transmitted concurrently with individual rates and outage probabilities, we assume a two queue / two server model.} with two buffers per user (HC and LC queues) as shown in Fig. \ref{fig:sys_model}. A similar two-queue model has been used in M2M uplink scheduling for managing delay-sensitive and delay-tolerant traffic \cite{kumar2016M2M}. The state evolution of the queues (i.e., the number of buffered data packets) at UE $i$ are given by \cite{chaccour2020risk}
\begin{equation}
    \GC{Q}_i(t+1) = \left[\GC{Q}_i(t) - \GC{L}_i(t)\right]^+ + \GC{A}_i(t), \quad c \in \{h,l\}.\label{queue_genC}
\end{equation}
In \eqref{queue_genC}, $\HC{A}_{i}(t)$ and $\LC{A}_{i}(t)$ represent the number of HC and LC packets generated at UE $i$ in time slot $t$, which are assumed to follow a Poisson distribution with arrival rates $\alpha_i\bar{a}_{\mathrm{t},i}$ and $(1-\alpha_i) \bar{a}_{\mathrm{t},i}$, respectively. Here, $\bar{a}_{\mathrm{t},i}$ is the total average rate of generated packets at UE $i$, and $\alpha_i \in [0,1]$ represents the ratio of packets classified as critical. 
$\HC{L}_{i}(t)$ and $\LC{L}_{i}(t)$ indicate the number of successfully decoded HC and LC data packets in time slot $t$.
In order to prevent buffer overflow, the transmission scheme should be designed in a way that all queues are stabilized. A queue is called stable, if
\begin{equation}
    \lim_{t\rightarrow \infty} \sup \frac{1}{t} \sum_{\tau=0}^{t-1} \mathbb{E} [Q(\tau)] < \infty.
\end{equation}
On a long-term basis, mean rate stability is achieved if the average departure rate is greater than the average arrival rate \cite{neely2010introduction}. 
Furthermore, the average waiting time of packets in the HC and LC buffer is, according to Little's law \cite{littlesLaw}, obtained as\looseness-1
\begin{equation}\label{delay_eq}
    \HC{\bar{\tau}}_i = \frac{\mathbb{E}\{\HC{Q}_i\}}{\alpha_i \bar{a}_{\mathrm{t},i}},\quad \LC{\bar{\tau}}_i = \frac{\mathbb{E}\{\LC{Q}_i\}}{(1-\alpha_i) \bar{a}_{\mathrm{t},i}}.
\end{equation}

\subsection{Problem Statement}
Our goal is to achieve resilient uplink communication in the presence of random channel blockages by mitigating packet loss and minimizing retransmissions. Specifically, we aim to reduce queuing delays for both HC and LC data, with a particular emphasis on ensuring timely delivery of HC traffic during blockage events. 
Numerous approaches for handling channel blockages have been studied, ranging from robustness-oriented methods such as multi-path transmission and multi-connectivity to link recovery mechanisms like handover and re-routing. Yet, the interplay and respective roles of these countermeasures within an integrated resilience framework remain under-explored. Coordinating them effectively to ensure service continuity, while avoiding resource overprovisioning and excessive overhead, continues to be a significant challenge. 

In fact, different mitigation techniques vary in effectiveness, response time, complexity, coordination costs, and resource consumption. 
Robustness through redundancy and diversity provides a first line of defense but becomes inefficient when failures are rare or prolonged. Conversely, adaptive strategies allow targeted recovery but often incur delays and overhead, which is unsuitable for mission-critical applications. The optimal deployment therefore depends on disruption characteristics, such as predictability and duration, as well as the heterogeneous service requirements of the application.

As temporary failures are inevitable in dynamic environments, resilience must also be complemented by criticality-aware schemes that prioritize time-sensitive data. Addressing these challenges requires a framework that integrates robustness and recovery mechanisms while balancing responsiveness and efficiency.
To this end, we propose an agile resilience framework composed of multiple stages that are activated sequentially in response to blockage events. Rather than focusing solely on individual strategies, our approach provides general principles for the coordinated activation and efficient use of multiple resilience techniques. The following section introduces this multi-stage framework and applies it to the uplink transmission scenario under consideration.

\section{Multi-Stage Resilience Strategy}\label{sec:resilience}
\begin{figure}[tb]
    \centering
    \includegraphics[scale=0.94]{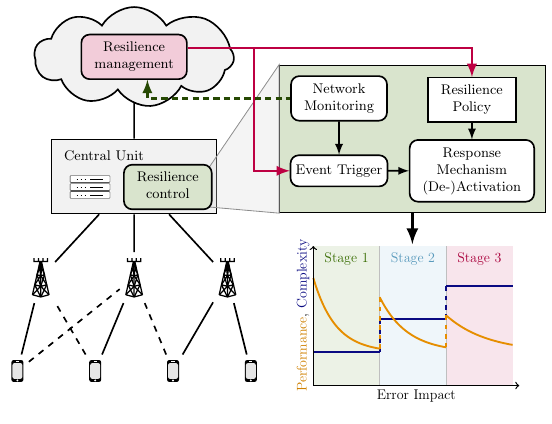}
    \caption{Resilience framework implemented on management and control plane. The plot illustrates the multi-stage strategy comprising different schemes with increasing complexity and error mitigation efficacy. As the impact of the error increases, switching to the next stage reduces the resulting performance degradation.}
    \label{fig:resilience_manager}
\end{figure}

The selection and activation of resilience mechanisms takes place on the network management and control plane. As illustrated in Fig. \ref{fig:resilience_manager}, a central control unit oversees near real-time adaptation of the communication scheme, based on predefined policies. We propose an event-based control algorithm that triggers specific resilience stages in response to failure detection and QoS monitoring. Cloud-based network management enables dynamic adjustment of triggering conditions and resilience policies to meet evolving requirements. 
In this work, we focus on the event-driven activation and deactivation of cooperative communication schemes to enable short-term resilience. Long-term policy adaptation is left for future work.

We implement the proposed agile resilience framework as a three-stage strategy tailored to dynamic link blockages in the uplink. While the framework provides general design guidelines for resilience and could be adapted to other types of disruptions, such as hardware/software failures or traffic overload, we focus in this work on its application to blockage-induced communication interruptions. As illustrated in Fig. \ref{fig:resilience_manager}, resilience mechanisms are activated progressively in response to detected events. 
This staged approach is conceptually similar to the ``gear-switching'' concept introduced in \cite{fettweis2021tactile}, where the modulation and coding scheme is adapted based on the available spectrum to balance data rate and energy consumption. Likewise, our framework escalates resilience mechanisms as needed, only engaging more resource-intensive measures when simpler methods are insufficient. Figure \ref{fig:resilience_manager} illustrates this progression, highlighting the tradeoff between operational cost and fault tolerance across stages.
The following outlines the three resilience stages in detail.

   \begin{figure*}[ht]
    \centering
    \includegraphics[scale=0.9]{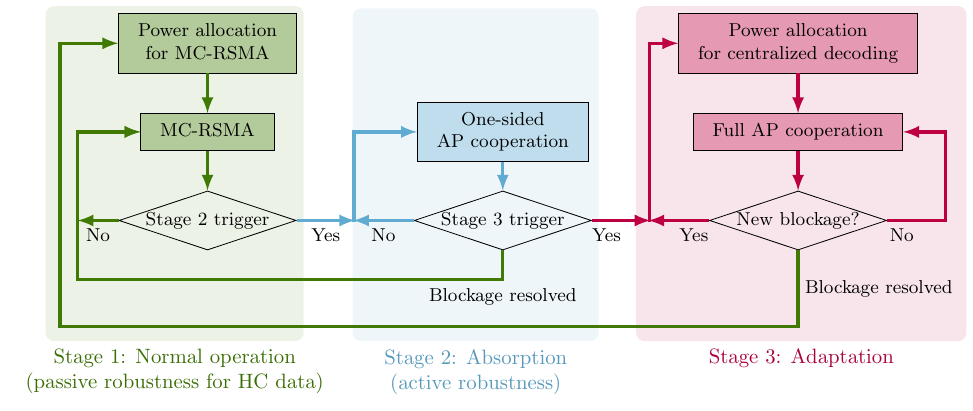}
    \caption{Event-triggered multi-stage resilience scheme}
    \label{fig:multi-stage}
\end{figure*}

\vspace{-10pt}
 \subsection*{Stage 1: Normal Operation with \emph{Passive Robustness} for High-Criticality Data} 
    This mode is active during regular operation, before any failure is detected. In anticipation of potential disruptions, such as LoS blockage, Stage 1 incorporates passive robustness strategies, e.g., redundant transmission paths and multi-link connectivity. These mechanisms are proactively employed to mitigate the impact of short-term impairments without requiring immediate reconfiguration. 
   However, permanent activation of such measures may lead to inefficient resource usage, especially during standard operation.
    To address this tradeoff and enhance resource efficiency, we propose a new criticality-aware design for robust communication. In this approach, enhanced robustness measures are selectively applied to critical data transmission, whereas non-critical data transmission is optimized for efficiency, thereby improving the overall system utilization without compromising the resilience of mission-critical communication.
    In the considered uplink scenario with dynamic blockages, this concept is realized through a novel \emph{mixed-criticality RSMA-based multi-connectivity scheme} to be applied in Stage 1 normal operation mode. Here, high-criticality data is decoded at multiple access points for increased robustness against link blockages.

   \vspace{-5pt}
    \subsection*{Stage 2: Absorption via \emph{Active Robustness}}
    If the passive robustness measures in Stage 1 are insufficient to maintain service functionality, the system transitions to Stage 2, namely \emph{active robustness}. This stage is activated to address errors more dynamically by deploying low-complexity adjustments in response to a detected failure. 
    However, rather than modifying the entire transmission scheme, Stage 2 responses remain confined locally and temporally, leaving other unaffected parts of the system untouched. This allows for efficient handling of short-term failures that affect only specific links or components, without introducing significant complexity and adaptation delays. Importantly, Stage 2 also supports the continuity of LC traffic during transient disruptions.
    
    In the considered scenario of uplink blockage, we propose \emph{one-sided AP cooperation} as the active robustness mechanism. 
    In this scheme, a blocked AP obtains the received signal from an unblocked cooperating AP to assist in decoding its intended messages. While this introduces a delay and quantization noise due to inter-AP signal exchange, the scheme deliberately avoids any reallocation of resources. Thereby it provides a lightweight and localized response that enhances resilience without requiring changes to the overall transmission strategy.

 \vspace{-5pt}
    \subsection*{Stage 3: Adaptation and Remediation}
    If a disruption persists beyond application-specific delay constraints or affects a broader portion of the system, the network transitions to Stage 3. This stage addresses errors that cannot be mitigated by predefined robustness mechanisms, particularly rare or prolonged failures. It involves higher-complexity adaptations that require broader coordination across network nodes and layers. While not suitable for frequent use due to increased overhead, these measures are essential when earlier stages fail to maintain service functionality.
    
    In this work, Stage 3 is realized through \emph{full AP cooperation with centralized decoding}, where joint processing of received signals from multiple APs enables improved decoding reliability during sustained blockages. This scheme also incorporates \emph{optimized transmit power allocation} at the cost of additional computational delay and coordination overhead.

   \vspace{2ex}These three stages are coordinated within an \emph{event-based} control framework as shown in Fig. \ref{fig:multi-stage}. The triggering conditions for Stage 2 and Stage 3 are defined later in Section \ref{sec:event_alg}. This agile approach enables both robustness and dynamic adaptation in a targeted and resource-efficient manner. 
The integration of \emph{criticality awareness} ensures that stringent delay requirements for high-priority traffic are met, even in the presence of failures, while maintaining support for throughput-oriented services. The specific transmission and decoding strategies employed in each stage are detailed in the following sections.

\section{Mixed-Criticality RSMA (Stage 1)}\label{sec:MC_RSMA}
In Stage 1, the UEs apply a criticality-aware multi-connectivity scheme, similar to the one we proposed in \cite{karacora2022ratesplitting}, to communicate their mixed-criticality data to the APs. While related to RSMA, the novelty of the proposed approach lies in its application as a flexible multi-/single-connectivity scheme designed to enhance blockage robustness.
In conventional RSMA, the transmit data is split into common and private message streams, which are jointly transmitted using superposition coding. By employing a successive interference cancellation (SIC) strategy, multiple receivers decode the common stream to mitigate interference, while the private stream is exclusively decoded by the UE's associated receiver and treated as noise by others. This strategy enables efficient interference management and enhances overall spectral efficiency \cite{mao2022rate}.
In contrast, our mixed-criticality uplink scheme (denoted as MC-RSMA \cite{karacora2022ratesplitting}) splits each user's data based on service criticality. HC data is transmitted such that it can be decoded by multiple APs, analogous to the common stream in RSMA, whereas LC data is decoded only by the designated AP, similar to private message handling. Thereby, this scheme enhances the reliability of HC data transmission in two ways: 
(1) The HC message is decoded first, whereas LC message decoding depends on the cancellation of the HC message, hence suffering from error propagation.
(2) Since the HC stream is decoded by multiple APs, this scheme inherently leverages spatial macro-diversity in the uplink to improve robustness against blockage-induced link failures.
Thus, if a link to one of the APs is weakened due to the temporary unavailability of the LoS path, the HC message stream can still be recovered with high probability via unaffected links. 
At Stage 1 of the resilience framework, which is deployed in normal operation mode, this selective redundancy improves robustness for critical data, while conserving resources for non-critical transmissions.
\begin{figure}
    \centering
    \includegraphics[width=0.94\linewidth]{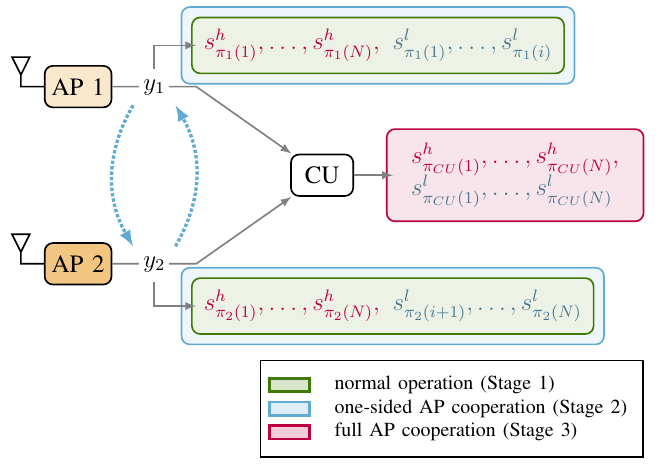}
    \caption{Decoding strategies of the schemes applied in each of the three proposed stages, where the permutation $\pi_j$ indicates the decoding order at AP $j$.}
    \label{fig:decoding}
\end{figure}

We apply superposition coding at the UE to jointly transmit both data streams. Thus, the transmit signal of UE $i$ is given by:
\begin{equation}
    {x}_i =  \sqrt{\HC{p}_{i}} \HC{s}_{i} + \sqrt{\LC{p}_{i}} \LC{s}_i,
\end{equation}
where $\HC{s}_i$ and $\LC{s}_i$ are the HC and LC messages, respectively, and $\HC{p}_i$ and $\LC{p}_i$ denote the corresponding transmit powers. 
The received signal at AP $j$ is obtained as
\begin{equation}
    y_j = \sum_{i=1}^N h_{ji} x_i + n_j,
\end{equation}
in which $n_j \sim \mathcal{CN}(0, \sigma_j^2)$ is additive white Gaussian noise.
The APs first decode the HC messages of all UEs by performing SIC, followed by LC message decoding of the associated users (see Fig. \ref{fig:decoding}). Here, the decoding order at AP $j$ is based on the expected channel gains and represented by the permutation $\pi_j$, where $\pi_j(i) < \pi_j(k)$ if $|\hat{h}_{ji}| > |\hat{h}_{jk}|$.
Consider binary decoding indicators for the HC and LC messages of the $i$-th UE, denoted as $\HC{\eta}_{ji}$ and $\LC{\eta}_i$. These indicators take the value of 1 if the corresponding message is successfully decoded (at AP $j$) and 0 in case of decoding failure, indicating an outage for the respective message. 
We further define the index set $\mathcal{I}^{int,h}_{ji} = \{k \in \mathcal{I}|\pi_j(k)>\pi_j(i)\}$, representing the HC messages that are supposed to be decoded after decoding $\HC{s}_i$ at AP $j$. Similarly, the set $\mathcal{I}^{int,l}_{ji} =  \mathcal{I}_{j'} \cup \{k \in \mathcal{I}_j|\pi_j(k)>\pi_j(i)\}$ captures the indices of LC messages not (yet) decoded when decoding $\LC{s}_i$ at AP $j$.
Then, the signal-to-interference-plus-noise ratio (SINR) for successive decoding at AP $j$ will be given by \eqref{gamma_c_ii} and \eqref{gamma_p_ii}.
\begin{figure*}[!t]
\begin{align}
    \HC{\Gamma}_{ji} &= \frac{|h_{ji}|^2 \HC{p}_i}{ \sum_{k \in \mathcal{I}}|h_{jk}|^2 \LC{p}_k + \sum_{m \in \mathcal{I}^{int,h}_{ji}} |h_{jm}|^2 \HC{p}_m + \sum_{\substack{n \in \mathcal{I} \setminus \mathcal{I}^{int,h}_{ji}, n\neq i}}(1-\HC{\eta}_{jn})|h_{jn}|^2 \HC{p}_n + \sigma_j^2}, \label{gamma_c_ii}\\
    \LC{\Gamma}_{ji} &=\frac{|h_{ji}|^2 \LC{p}_i}{ \sum_{k \in \mathcal{I}^{int,l}_{ji}}|h_{jk}|^2 \LC{p}_k + \sum_{m \in \mathcal{I}}(1-\HC{\eta}_{jm})|h_{jm}|^2 \HC{p}_m + \sum_{n\neq i,n \in \mathcal{I} \setminus \mathcal{I}^{int,l}_{ji}}(1-\LC{\eta}_{jn})|h_{jn}|^2 \LC{p}_n + \sigma_j^2}. \label{gamma_p_ii}
\end{align}
\vspace{-2pt}
\hrule\vspace{-10pt}
\end{figure*}
For received SINR $\Gamma$ and bandwidth $B$, the achievable rate is $R=B \log_2(1 + \Gamma)$. 
When none of the links is blocked, each AP should successively decode all HC messages and the LC messages of its associated UEs. However, in the case of a link blockage, we leverage multi-connectivity for the HC message and consequently adjust the coding rate by relaxing the multi-connectivity constraint that enforces HC message decoding at both APs. Thus, for user $i \in \mathcal{I}_j,~j'\neq j$, the coding rates for the HC stream are determined based on the detected LoS/NLoS channel states given by $\hat{\beta}$ as follows:
\begin{equation}
    \HC{\hat{R}}_i = \begin{cases}B \log_2\left(1 + \min\{\HC{\hat{\Gamma}}_{ji}, \HC{\hat{\Gamma}}_{j'i}\}\right), & \text{if } \hat{\beta}_{ji}=\hat{\beta}_{j'i}=1, \\ B \log_2\left(1 + \HC{\hat{\Gamma}}_{j'i}\right),  & \text{if } \hat{\beta}_{ji}=0,~ \hat{\beta}_{j'i}=1,\\  B \log_2\left(1 + \HC{\hat{\Gamma}}_{ji}\right), & \text{if } \hat{\beta}_{j'i}=0. \end{cases}
\end{equation}
The coding rate for the LC message of UE $i \in \mathcal{I}_j$ is given by $\LC{\hat{R}}_i = B \log_2\left(1 + \LC{\hat{\Gamma}}_{ji}\right)$. 
Here, the expected SINR expressions are obtained as
\begin{align}
 \hspace{-1.5pt}\HC{\hat{\Gamma}}_{ji} &= \frac{|\hat{h}_{ji}|^2 \HC{p}_i}{\displaystyle \sum_{k \in \mathcal{I}}|\hat{h}_{jk}|^2 \LC{p}_k + \hspace{-3pt}\sum_{m \in \mathcal{I}^{int,h}_{ji}}\hspace{-3pt} |\hat{h}_{jm}|^2 \HC{p}_m + \sigma_j^2},  ~~i \in \mathcal{I},\label{gamma_hat_c_ii}\\
    \hspace{-1.5pt}\LC{\hat{\Gamma}}_{ji} &=\frac{|\hat{h}_{ji}|^2 \LC{p}_i}{\displaystyle \hspace{-4pt} \sum_{k \in \mathcal{I}^{int,l}_{ji}}\hspace{-3pt}|\hat{h}_{jk}|^2 \LC{p}_k + \hspace{-2pt} \sum_{m \in \mathcal{I}}(1\hspace{-1pt}-\hspace{-1pt}\hat{\beta}_{jm})|\hat{h}_{jm}|^2 \HC{p}_m  + \sigma_j^2},~ i \in \mathcal{I}_j. \label{gamma_hat_p_ii}
\end{align}
Note that when blockage of an interference link $h_{jm},~m\not\in \mathcal{I}_j$, is detected, the HC message of UE $m$ will no longer be decoded for interference cancellation at AP $j$, but instead it will be treated as noise. Hence, it is represented by the second interference term in \eqref{gamma_hat_p_ii} and the coding rates are adjusted accordingly in the time slot following the blockage detection. This, in turn, leads to a decreased LC rate for UE $i$ in favor of a higher HC rate for UE $m$.
Thus, we have
\begin{equation}
    \GC{\eta}_{ji} = \begin{cases} 1, & \text{if $\GC{R}_{ji} \geq \GC{\hat{R}}_i$}, \\ 0, & \text{otherwise,}\end{cases},\quad c\in \{h,l\}.
\end{equation}
Given that the HC data exhibits higher reliability by being decoded at multiple APs, the successfully decoded packets departing the UEs buffers are obtained as   
\begin{align}
    \HC{L}_{i} &= \max\{\HC{\eta}_{ji}, \HC{\eta}_{j'i}\}\cdot \HC{\hat{R}}_{i},\\
    \LC{L}_{i} &= \LC{\eta}_j \LC{\hat{R}}_i,\quad i \in \mathcal{I}_j.
\end{align}

Our goal is to optimize power allocation for HC and LC message streams at each user, ensuring queue stability across the system while minimizing computational overhead and user coordination. In our proposed MC-RSMA scheme, power allocation targets mean queue stability by ensuring that each buffer's average service rate exceeds the average packet arrival rate. This approach minimizes the need for frequent power allocation optimization, thus reducing computational overhead and centralized coordination.
To maintain fairness among the UEs, we consider a max-min optimization problem that leads to equal utilization of all queues. The utilization (i.e., traffic intensity) of a queuing system is defined as the arrival rate divided by the service rate.
For tractability, we define the optimization variable $\delta_i = \frac{\mathbb{E}[L_i]}{\mathbb{E}[A_i]}$ as the inverse utilization. 
Then, the power allocation problem for the MC-RSMA scheme is formulated as follows:
\begin{subequations}
\label{opt1}
\begin{align}
\max_{\vec{p}, \vec{R}, \vec{\delta}} ~& \underset{i}{\min}\{\HC{\delta}_i, \LC{\delta}_i\} \tag{\theparentequation}\vspace{-2mm}\\
\text{s.t.}~~ & \mathbb{E}\left[\max\{\HC{\eta}_{ji}, \HC{\eta}_{j'i}\} \right] \frac{T}{M} \HC{R}_i \geq \alpha_i\bar{a}_{\mathrm{t},i} \HC{\delta}_i, \hspace{-3cm} & &\label{HC_stability_constr}\\
 & \mathbb{E} \left[\LC{\eta}_i\right]\frac{T}{M} \LC{R}_i \geq (1-\alpha_i)\bar{a}_{\mathrm{t},i} \LC{\delta}_{i}, & &\label{LC_stability_constr} \\
&\HC{r}_{ji} \leq B \log_2\left( 1 + \HC{\bar{\Gamma}}_{ji}\right), &  &\label{c_rate_constr}\\
&\HC{R}_i \leq \min\{\HC{r}_{ji}, \HC{r}_{j'i}\}, & j' \neq j, &\label{RSMA_constr}\\
&\LC{R}_{i} \leq B \log_2\left(1 + \LC{\bar{\Gamma}}_{ji}\right), & i \in \mathcal{I}_j, & \label{p_rate_constr}\\
 & \HC{p}_i + \LC{p}_i \leq P^\mathrm{max}_i, & i \in \mathcal{I}.& \label{power_constr}
\end{align}
\end{subequations}
Together with the constraints \eqref{HC_stability_constr} and \eqref{LC_stability_constr}, the objective is to maximize the ratio of average service rate and the average arrival rate (i.e., the inverse utilization) of the queue which is most prone to becoming instable, thereby ensuring fairness.
Note that mean stability holds for UE $i$ as long as $\HC{\delta}_{i}>1$ and $\LC{\delta}_{i} >1$. Apart from that, \eqref{c_rate_constr}, \eqref{RSMA_constr}, and \eqref{p_rate_constr} are the rate constraints for the HC and LC data rates, respectively. Here, the SINR expressions $\HC{\bar{\Gamma}}_{ji}$ and $\LC{\bar{\Gamma}}_{ji}$ are obtained based on the average channel gains, i.e., from \eqref{gamma_c_ii} -- \eqref{gamma_p_ii} by replacing $|h_{ji}|^2$ with $\mathrm{PL}_{ji},~ i\in \mathcal{I}$.
Moreover, \eqref{power_constr} is the power constraint for each UE.

\YK{The formulation in \eqref{opt1} significantly differs from classical RSMA power allocation problems, which typically aim to optimize the sum of common and private rates for physical-layer interference management. In contrast, our approach adopts a cross-layer perspective by ensuring queue stability for HC and LC streams, which are mapped to the common and private layers, respectively. Consequently, the power allocation must meet these criticality-aware rate requirements, rather than merely maximizing total throughput. Moreover, the proposed scheme functions as a hybrid single-/multi-connectivity mechanism, shifting the focus from interference mitigation to resilient and efficient support for mixed-criticality services under dynamic link conditions.}

We first approximate the expectation expressions in the constraints \eqref{HC_stability_constr} and \eqref{LC_stability_constr} by assuming that decoding fails for the entire time slot if the LoS component of the desired channel is blocked.
Based on the decoding order $\pi_j$ at AP $j$ and considering error propagation in case of undetected blockages, we obtain the following approximate success probabilities at AP $j$:
\begin{align*}
    P(\HC{\eta}_{ji}=1) &\approx (1-p_b) \cdot \left(1- (1-p_b) (1-e^{-\kappa_B T})\right)^{\pi_j(i)-1} ,\\
    P(\LC{\eta}_i =1) &\approx(1-p_b)\cdot \left(1 - (1-p_b)(1-e^{-\kappa_B T})\right)^{N-1}.
\end{align*}
Since the HC messages are delivered successfully if decoded at any AP, the success probability of $\HC{s}_i$ is given as
\begin{equation*}
 P\left(\max\{\HC{\eta}_{ji}, \HC{\eta}_{j'i}\}=1\right) =  P(\HC{\eta}_{ji}=1) + P(\HC{\eta}_{ji}=0) P(\HC{\eta}_{j'i}=1).
\end{equation*}
 Problem \eqref{opt1} is not convex due to the non-convexity of the constraints \eqref{c_rate_constr} and \eqref{p_rate_constr}. However, we apply a successive convex approximation method together with fractional programming involving a quadratic transform \cite{shen2018fractional}. The convex approximation is derived following similar steps as in \cite{karacora2024intermittency}, which is omitted here for brevity. Finally, a locally optimal power allocation can be computed iteratively by means of a convex optimization solver such as CVX \cite{cvx}.
 
 In essence, Stage 1 improves reliability for HC data through continuous multi-connectivity, allowing the system to tolerate short-term disruptions without requiring failure detection or reconfiguration. To preserve efficiency, this protection is limited to HC data, while LC transmissions are more vulnerable to blockage-induced decoding failures, leading to queue buildup. In response, Stages 2 and 3 aim to prevent buffer overflow and maintain service continuity through AP cooperation.

\section{Cooperation Schemes}\label{sec:coop}
\subsection{One-Sided AP Cooperation (Stage 2)}\label{sec:one_sided_coop}
LoS blockages inevitably lead to performance drops for LC traffic, risking service degradation. To counteract such disruptions without altering the transmission configuration, we propose an on-demand, one-sided AP cooperation scheme.
That is, in Stage 2, if AP $j$ is affected by blockage in time slot $t$, AP $j'$ shares its received signal $y_{j'}(t)$ to help AP $j$ decode its desired messages. Then, in time slot $t+1$, AP $j$ performs successive decoding of the messages $\{\HC{s}_{i}\vert i\in \mathcal{I}\}$ and $\{\LC{s}_{i}\vert i\in\mathcal{I}_j\}$ based on the combined receive signal from both APs (see Fig. \ref{fig:decoding}). 
Meanwhile, new data packets can be transmitted in time slot $t+1$ in parallel to the cooperative decoding process.

More precisely, the decoding process with cooperation at AP $j$ is as follows:
After initiating the cooperative decoding in time slot $t$, AP $j$ acquires the receive signal from AP $j'$ in time slot $t+1$. AP $j$ then decodes its desired messages using maximum ratio combining (MRC) and SIC similar to the cooperative RSMA scheme detailed in \cite{abbasi2022transmission}. First, MRC is applied to combine the received signals $y_j(t)$ and $y_{j'}(t)$ for detection of the HC message $\HC{s}_i$ with $\pi_j(i)=1$. Upon successful decoding and interference removal from both signals via SIC, MRC is again employed to decode the subsequent HC message. This iterative process continues, enabling successive decoding of all HC messages and intended LC messages via MRC and SIC. 
The achievable rates at AP $j$ using this cooperative decoding procedure are determined as follows:
\begin{equation}
    \GC{R}_{\mathrm{coop},ji} = B \log_2 \left(1 + \GC{\Gamma}_{ji} + \GC{\tilde{\Gamma}}_{j'i} \right), \quad c\in\{h,l\}.\label{R_S2_ci}
\end{equation}
Here, $\HC{\Gamma}_{ji}$ and $\LC{\Gamma}_{ji}$ are given in \eqref{gamma_c_ii} and \eqref{gamma_p_ii}. 
Suppose that the signal of the cooperative AP received in time slot $t$ is quantized before being forwarded to the other AP in the following time slot, whereby quantization noise is introduced.
Assuming dithered quantization \cite{gray1993quantizer} of $y_{j'}$, the quantization noise is independent of the quantized signal, and adopting rate distortion theory \cite{cover1999elements, fritzsche2013backhaul} the quantization noise variance is obtained as $\sigma_{\mathrm{q},j'}^2 = 2^{-N_q} \sigma_{y_{j'}}^2$. Here, $N_q$ is the number of quantization bits and the variance of the receive signal is given by $\sigma_{y_{j'}}^2= \sum_{i \in \mathcal{I}}P_{i}^\mathrm{max}\mathrm{PL}_{j'i} + \sigma_{j'}^2$.
Thus, the overall noise variance is $\tilde{\sigma}_{j'}^2 = \sigma_{j'}^2+\sigma_{\mathrm{q},j'}^2$, and the corresponding SINR expressions for $\HC{\tilde{\Gamma}}_{j'i}$ and $\LC{\tilde{\Gamma}}_{j'i}$ are obtained based on \eqref{gamma_c_ii} and \eqref{gamma_p_ii} by replacing $\sigma_{j'}^2$ by $\tilde{\sigma}_{j'}^2$, respectively.
The coding rates are determined based on \eqref{gamma_hat_c_ii}-\eqref{gamma_hat_p_ii} in an analogous manner, while ensuring HC message decoding at both APs.

In summary, Stage 2 provides a reactive, transitional solution to enhance resilience against blockage. Its on-demand deployment minimizes additional cooperation costs, and by keeping transmit powers unchanged while adjusting only the coding rate based on real-time channel measurements, this strategy enables a quick response without optimization delays.
However, for ongoing disruptions, one-sided cooperation may fall short of performance needs, necessitating an adaptive solution as provided in Stage 3.

\subsection{Full AP Cooperation / Central Decoding (Stage 3)}\label{sec:full_coop}
In Stage 3, we consider full AP cooperation with centralized decoding, where the APs forward their received signals to the CU for joint processing (see Fig. \ref{fig:decoding}). Thereby, spatial macro-diversity is exploited for both the HC and LC data and the achievable rate increases compared to separate decoding at each AP. Note that this centralized decoding approach comes with an increase in latency depending on the fronthaul capacity. This is why for latency-sensitive applications, decentralized decoding at the APs is the preferred strategy, while joint decoding at the CU can be temporarily applied in order to maintain service functionality during a link blockage phase.

Again, we assume that signals received by the APs in time slot $t$ are quantized before being forwarded to the CU via the fronthaul links during the following time slot. Hence, the total received signal at the CU in time slot $t+1$ is given as
\begin{equation}
    \vec{y}_\mathrm{CU}(t+1) = \begin{pmatrix} y_1(t) \\ y_2(t) \end{pmatrix} = \sum_{i\in\mathcal{I}}\vec{h}_i(t) x_i(t) + \vec{n}(t),
\end{equation}
where $\vec{h}_i=[h_{1i}, h_{2i}]^T$,  and $\vec{n}\sim \mathcal{CN}(0, \diag \{\tilde{\sigma}_1^2,\tilde{\sigma}_2^2\})$.

The CU successively decodes the data streams based on the joint receive signal.
\MU{We assume that the decoding order is again determined based on the channel gain, so that $\pi_\mathrm{CU}(i) < \pi_\mathrm{CU}(k)$ if  $||\vec{h}_i|| > ||\vec{h}_k||$. Defining the index set $\mathcal{I}^{int}_{\mathrm{CU},i}=\{k \in \mathcal{I} ~|~ \pi_\mathrm{CU}(k) > \pi_\mathrm{CU}(i)\}$,} the achievable SINR expressions with full AP cooperation are given in \eqref{R_S3_ci}--\eqref{R_S3_pi} (at the top of the next page).
\begin{figure*}[htb]
    \begin{align}
  \HC{\mat{\Gamma}}_{\mathrm{CU},i} &= \vec{h}_i \vec{h}_i^H \HC{p}_i \left.\Biggl( \sum_{k\in\mathrm{I}} \vec{h}_k \vec{h}_k^H \LC{p}_k  +\hspace{-3pt} \sum_{m\in\mathcal{I}^{int}_{\mathrm{CU},i}} \hspace{-3pt}\vec{h}_m \vec{h}_m^H \HC{p}_m \right. \hspace{-0em}\left.+ \hspace{-3pt}\sum_{\substack{n\in\mathcal{I}\setminus\mathcal{I}^{int}_{\mathrm{CU},i},n\neq i}}\hspace{-3pt} (1-\HC{\eta}_n)\vec{h}_n \vec{h}_n^H\HC{p}_n + \diag\{\tilde{\sigma}_1^2,\tilde{\sigma}_2^2\} \right.\Biggr)^{-1}, \label{R_S3_ci}\vspace{-2pt}\\ 
    \LC{\mat{\Gamma}}_{\mathrm{CU},i} &= \vec{h}_i \vec{h}_i^H \LC{p}_i \left.\Biggl(\sum_{k\in\mathcal{I}^{int}_{\mathrm{CU},i}} \hspace{-2pt}\vec{h}_k \vec{h}_k^H \LC{p}_k + \sum_{m\in\mathcal{I}} (1-\HC{\eta}_{m})\vec{h}_m \vec{h}_m^H \HC{p}_m \right.+ \left.\hspace{-3pt}\sum_{\substack{n\in\mathcal{I}\setminus\mathcal{I}^{int}_{\mathrm{CU},i},\\n\neq i}} \hspace{-3pt}\vec{h}_n \vec{h}_n^H (1-\LC{\eta}_n)\HC{p}_n  +\diag\{\tilde{\sigma}_1^2,\tilde{\sigma}_2^2\} \right.\Biggr)^{-1}.\label{R_S3_pi}
\end{align}
\vspace{-5pt}
\hrule
\end{figure*}
Thus, the achievable rates are obtained by $R= B \log_2\left( \det(\mat{I}+\mat{\Gamma})\right)$. Likewise, the coding rates are determined as $\hat{R}= B \log_2\left( \det(\mat{I}+\mat{\hat{\Gamma}})\right)$, in which the estimated SINR is obtained from \eqref{R_S3_ci}-\eqref{R_S3_pi} by replacing $\vec{h}$ by $\vec{\hat{h}}$ and setting all $\eta$ to 1.

At Stage 3, the optimal power allocation under full AP cooperation is calculated based on the expected path loss including the detected LoS blockage. Here, resources are reallocated for mean queue stability whenever a newly blocked or unblocked link is detected.
Similar to problem \eqref{opt1}, we formulate the following optimization problem to obtain the optimal power allocation for the central decoding scheme:
\begin{subequations}
\label{opt_S3}
\begin{align}
\max_{\vec{p}, \vec{R}, \vec{\delta}} ~& \underset{i}{\min}\{\HC{\delta}_i, \LC{\delta}_i\} \tag{\theparentequation}\vspace{-2mm}\\
\text{s.t.}~~ & \mathbb{E}\left[\HC{\eta}_i\right] \frac{T}{M} \HC{R}_{\mathrm{CU},i} \geq \alpha_i\bar{a}_{\mathrm{t},i} \HC{\delta}_{i}, & \label{HC_stability_constr_S3}\\
 & \mathbb{E} \left[\LC{\eta}_i\right]\frac{T}{M} \LC{R}_{\mathrm{CU},i} \geq (1-\alpha_i)\bar{a}_{\mathrm{t},i} \LC{\delta}_{i}, & \label{LC_stability_constr_S3} \\
&\HC{R}_{\mathrm{CU},i} \leq B \log_2\left( \det(\mat{I} + \HC{\mat{\bar{\Gamma}}}_{\mathrm{CU},i})\right), &\label{R_c_constr_S3}\\
&\LC{R}_{\mathrm{CU},i} \leq B \log_2\left(\det(\mat{I} + \LC{\mat{\bar{\Gamma}}}_{\mathrm{CU},i})\right), &\label{R_p_constr_S3}\\
 & \HC{p}_i + \LC{p}_i \leq P^\mathrm{max}_i, & i \in \mathcal{I}.
\end{align}
\end{subequations}

In order to solve \eqref{opt_S3}, the constraints \eqref{HC_stability_constr_S3} and \eqref{LC_stability_constr_S3} are approximated as follows. In Stage 3, the power allocation is adjusted whenever blocking or unblocking of a link is detected. Thus, we assume that decoding fails when any of the links LoS/NLoS state changes. 
 Hence, we have
\begin{equation}
    \mathbb{E}\left[\HC{\eta}_i\right] =  \mathbb{E}\left[\LC{\eta}_i\right] = \left(1-e^{-\kappa_\mathrm{B}T}\right)^n \left(1-e^{-\mu_\mathrm{B}T}\right)^{(N-n)},
\end{equation}
where $n$ is the number of non-blocked LoS links.
While problem \eqref{opt_S3} is again non-convex, it can be efficiently solved using SCA and fractional programming in a similar way as with \eqref{opt1}.

Hence, while Stage 1 provides passive robustness for HC data, and Stage 2 enables a rapid response to blockages also partly mitigating LC transmission outages, Stage 3 \emph{adapts} to LoS disruption with optimized power allocation and centralized decoding. These three proposed schemes are embedded into an event-based resilience framework next.

\section{Coordinated Deployment of Resilience Stages}\label{sec:event_alg}

To ensure efficient and adaptive operation, the three resilience stages must be coordinated based on their cost and their benefit under current network conditions. This section first analyzes the costs introduced by each stage. Then the event-driven, sequential activation scheme (as shown in Fig. \ref{fig:multi-stage}) is introduced.

\subsection{Cost Factors per Resilience Stage}
Each resilience stage introduces a different level of complexity and coordination overhead, which must be balanced against its performance benefits.
    While higher cooperation levels more effectively counteract link blockages, they also require additional signaling, computational, and fronthaul costs. 
    A detailed cost assessment of the three proposed transmission schemes lies beyond this work's scope, as comparing hardware, software, power, and fronthaul costs varies by system, resource availability, and requirements. Instead, we analyze conceptual trends of operational costs to provide qualitative insights into resilience versus cost tradeoffs.
    More precisely, we consider three cost factors: 
    \begin{itemize}
         \item $\rho_\mathrm{MC} \in [0,1]$ indicates the average usage of multi-connectivity per user. It is defined as the fraction of time slots, where a {HC} message from a user must be requested from another {AP}.
        \item $\rho_{\mathrm{coop},j} \in [0,1]$ represents the fronthaul use of {AP} $j$ due to cooperative signal exchange. It is defined as the fraction of time slots, where {AP} $j$ forwards its received signal via the fronthaul link.
        \item $\rho_\mathrm{opt} \in [0,1]$ denotes the frequency of power re-optimization, defined as the ratio of algorithm runs to the total number of time slots.
    \end{itemize}
    \begin{table}[tb]
\renewcommand{\arraystretch}{1.1}
    \centering
    \vspace{3pt}
    \caption{Comparison of cost factors for the three resilience stages.}
    \rev{
    \begin{tabular}{|c|ccc|}
    \hline
        \textit{Cost factors} & Stage 1 & Stage 2 & Stage 3 \\ \hline
        Multi-connectivity & $\rho_\mathrm{MC} \geq 0$ & $\rho_\mathrm{MC} = 0$ & $\rho_\mathrm{MC} = 0$\\ 
        Cooperation & $\rho_{\mathrm{coop},j} =0$ & $\rho_{\mathrm{coop},j} \geq 0$ & $\rho_{\mathrm{coop},j} = 1$\\ 
        Optimization & $\rho_\mathrm{opt} =0$ &  $\rho_\mathrm{opt} =0$ &  $\rho_\mathrm{opt} \geq0$\\ \hline
    \end{tabular}}
    \vspace{-8pt}
    \label{tab:complexity}
\end{table}
Table \ref{tab:complexity} summarizes the differences between the three schemes in terms of cost factors $\rho$.
In Stage 1, only minimal overhead arises from exploiting multi-connectivity when necessary, whereas Stage 2 introduces fronthaul traffic due to (one-sided) cooperative signal exchange. Stage 3 incurs the most overhead, with continuous fronthaul signal forwarding and occasional resource allocation adjustments based on changing channel conditions.

As a consequence of the increasing operational costs associated with more sophisticated and effective cooperation strategies, an event-based deployment emerges as a solution to balance costs and resilience. This approach activates complex mitigation techniques only when necessary to address severe performance declines, while simpler methods are preferred to overcome short outages and ensure critical service stability. The event-driven algorithm for dynamic activation/deactivation of the stages is presented next.

\subsection{Event-Driven Stage Activation}
In normal operation, the proposed MC-RSMA scheme (detailed in Sec. \ref{sec:MC_RSMA}) provides passive robustness for HC traffic. When a LoS blockage occurs that leads to buffer congestion, Stage 2 is triggered, and one-sided AP cooperation (as described in Sec. \ref{sec:one_sided_coop}) is applied as needed to facilitate message decoding. 
If this strategy fails to restore service stability, Stage 3, the adaptation mechanism, is activated (see Sec. \ref{sec:full_coop}). In this stage, we switch to central decoding of all UE's messages, and the transmit power allocation is optimized accordingly. The system returns to normal operation mode once the blockage is resolved.\looseness-1

In the considered scenario, rate and delay performance are reflected in the UE buffer backlog. Hence, to ensure timely packet delivery, the goal is to keep queues short.
To this end, we define thresholds $\HC{Q}_{\mathrm{max},i}, ~\LC{Q}_{\mathrm{max},i}$ for the amount of packets waiting in the buffer of UE $i$. These thresholds trigger a transition to different resilience measures as follows:

\paragraph{Stage 2 Trigger} If the HC and/or LC queue backlogs of UE $i$ exceed $Q^{h/l}_{\mathrm{max},i}$ \emph{due to a detected link blockage}, Stage 2 with one-sided AP cooperation is activated at AP $j$, $i\in\mathcal{I}_j$ i.e., AP $j$ requests the received signal of AP $j'$, $j'\neq j$ to support the decoding of its intended messages. 

\paragraph{Stage 3 Trigger} If the system is already in Stage 2 and any buffer exceeds the threshold for $n_e$ consecutive time slots, the scheme escalates to Stage 3.
The CU takes over central decoding of all messages and initiates power allocation optimization as described in Section \ref{sec:full_coop}.
The parameter $n_e$ reflects the delay induced by optimizing power allocation. Thereby, Stage 3 is activated only if Stage 2 fails to reduce the queue backlog below the threshold after multiple time slots. When the blockage state changes in Stage 3, transmit power is re-optimized with a delay of $n_e$ slots. 
Aside from a single-slot detection delay, triggers are assumed to take immediate effect in the next time slot.
The system reverts to Stage 1 when all blocked links are recovered.

\section{Simulation Results and Analysis}\label{sec:results}
We evaluate the performance of our proposed three-stage resilience scheme via numerical simulations.
We consider $N$ users randomly located within an area of size $150$ m $\times$ $150$ m, with two APs located at the corners, i.e., at coordinates $(0,0)$ and $(150,150)$, respectively. The user devices are uniformly distributed with a minimum distance of $50$ m to the APs.
The path loss is modeled as $\mathrm{PL}_{ij} = 128.1 + 37.6 \log_{10}(d_{ij}) $ dB, where the distance is in km \cite{yang2019uplinkRSMA}.
The Rician $K$-factor is assumed to be $K=20$. Other simulation parameters are given in Table \ref{tab:sim_param}.

\begin{table}[tb]
    \centering
    \vspace{3pt}
    \caption{Simulation parameters}
    \begin{tabular}{|c|c|}
    \hline
        Number of UEs $N$  &  $2$ / $8$ \\ \hline
       Transmit power/user $P^\mathrm{max}$  &  $10$ dBm\\ \hline
       Noise spectral density $N_0$ & $-174$ dBm/Hz\\ \hline
       Bandwidth $B$ & $4$ MHz / $20$ MHz \\ \hline
       Packet size $M$ & $1$ KBit/packet\\ \hline
       Time slot duration $T$ & $10$ ms \\ \hline
       Packet arrival rate $\bar{a}_\mathrm{t}$ & 10 Mbps \\ \hline
       HC packets fraction $\alpha$ & $0.5$\\ \hline
       Quantization bits $N_q$ & 10\\ \hline
       Queue thresholds $[\HC{Q}_{\mathrm{max}},~\LC{Q}_\mathrm{max}]$ & $[300, 1000]$ packets \\ \hline
       Stage 3 optimization delay $n_e$ & 10 slots \\ \hline
    \end{tabular} \vspace{-8pt} 
    \label{tab:sim_param}
\end{table}

\begin{figure}
    \centering
    \includegraphics[]{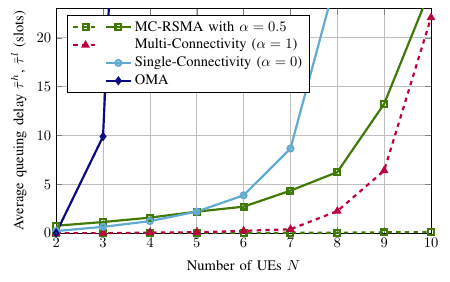}
    \caption{Average queuing delay of the worst-case user as a function of the number of users. Shown are the HC (dashed) and LC delays (solid) for the proposed MC-RSMA with $\alpha=0.5$, and baseline results for single-connectivity ($\alpha=0$), full multi-connectivity ($\alpha=1$), and single-connectivity OMA. Random blockages occur with $p_B=0.05$ and $\frac{1}{\mu_B}=100$ ms.}
    \label{fig:scalability}
\end{figure}

In Fig. \ref{fig:scalability}, we first evaluate the proposed MC-RSMA in isolation to highlight the benefits of the criticality-aware hybrid single-/multi-connectivity scheme. We plot the average worst user queuing delay, $\HC{\bar{\tau}}$ and $\LC{\bar{\tau}}$ as obtained from \eqref{delay_eq}, as a function of the number of UEs. Since our optimization is fairness-oriented, we evaluate the performance of the worst user in each Monte-Carlo run. As baselines, we consider single connectivity ($\alpha=0$) and full multi-connectivity ($\alpha=1$), which represent special cases of MC-RSMA. In these three cases, UEs share the spectrum non-orthogonally, whereas an additional baseline is orthogonal multiple access (OMA) with single-connectivity.  

This OMA scheme can only serve up to three users before queues become instable. With non-orthogonal transmission and single-connectivity ($\alpha=0$), the queuing delay is small when only a few UEs are active, but it grows rapidly with $N$. Here, blockages cause outages that cannot be compensated, and with more users competing for resources, queues are emptied more slowly after a blockage. Full multi-connectivity provides robustness to mitigate these outages, which keeps the delay much lower than in the single-connectivity case. However, because robustness is provided for all data, resource consumption is much higher, and the delay again increases sharply as $N$ grows.

MC-RSMA ($\alpha=0.5$) achieves a favorable balance by giving only HC data multi-connectivity while LC data uses single connectivity. This ensures that the HC delay stays close to zero even for large $N$. The LC delay grows with the number of users, yet more slowly than in the single-connectivity case. 
Notably, for $N<5$, single-connectivity achieves slightly lower LC delay, since no resources are spent on multi-connectivity. However, as $N$ increases, MC-RSMA outperforms the single-connectivity baseline. In this regime, the advantage of RSMA becomes evident: The LC data in MC-RSMA is transmitted with lower power and decoded after HC interference cancellation, enabling more resource-efficient rate gains and mitigating delay growth.

\begin{figure}[tb]
    \centering
    \includegraphics[width=.9\linewidth]{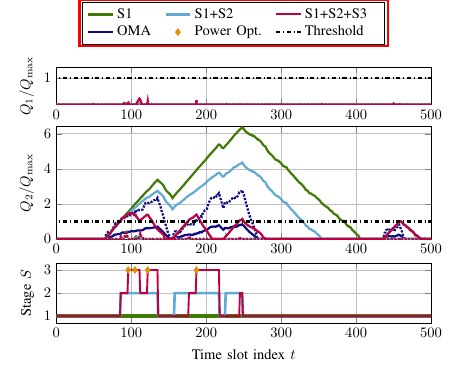}
    \caption{Evolution of normalized queue states over time of user 1 (top) and user 2 (middle), and activated resilience stages $S(t)$ (bottom) for random blockages with $\frac{1}{\mu_\mathrm{B}}=300$ ms and link blockage probability $p_b=0.05$. Solid lines and dotted lines represent the LC and HC queue states, normalized by their respective threshold $\LC{Q}_\mathrm{max}=1000$ and $\HC{Q}_\mathrm{max}=300$, respectively.}
    \label{fig:q_time}
\end{figure}

Next, we analyze the interplay of the three resilience stages within the event-driven framework.
Fig. \ref{fig:q_time} shows the queue state evolution for HC and LC traffic over 500 time slots ($T=10$ ms) for two users. The number of buffered packets is normalized by their respective thresholds $\HC{Q}_\mathrm{max}$ and $\LC{Q}_\mathrm{max}$. We analyze three resilience policies: Stage 1 only, Stages 1 and 2 combined, and the full three-stage algorithm based on the proposed event-driven approach in Sec. \ref{sec:event_alg}. Additionally, performance is compared with an OMA baseline scheme, which avoids interference between users, but operates a single data stream without service differentiation. Note that while only a single queue is simulated for OMA, the queue states in Fig. \ref{fig:q_time} are normalized considering the different queue thresholds $\HC{Q}_\mathrm{max}$ (dotted line) and $\LC{Q}_\mathrm{max}$ (solid line) as well as the HC fraction $\alpha$ for fair comparison.

During the simulation, multiple blockages intermittently affect the links $h_{22}$ and $h_{12}$ within the interval $t \in [65, 250]$, leading to data accumulation in the buffers of User 2.
Although in the 2-user scenario, RSMA's benefits over orthogonal schemes are less pronounced than in larger networks with more interference, the OMA scheme suffers severe HC threshold violations during these blockages. 
In contrast, all variations of the proposed framework prevent HC queue violations, leveraging Stage 1's multi-connectivity to provide robustness against temporary link failures. This underscores the need for a communication scheme that differentiates data criticality and provides QoS guarantees tailored to the specific requirements.
While Stage 1 prevents HC queue violations, it does not fully address LC queue congestion, leading to a prolonged threshold violation lasting around 300 time slots (3 seconds). Introducing Stage 2 (i.e., one-sided cooperation) alleviates some congestion but still fails to fully stabilize the LC queue. Since both Stage 1 and Stage 2 lack power optimization and instead rely on the solution from \eqref{opt1}, clearing the LC buffer takes considerable time even after blockages are resolved.
Only the full three-stage algorithm effectively stabilizes the LC queue during extended blockages. After a blockage of $h_{22}$ (starting at $t=66$), Stage 2 triggers, followed by Stage 3 after a 10-slot delay, optimizing power allocation and reducing LC queue buildup.
 When $h_{12}$ also becomes blocked, power allocation is further adjusted at $t=105$. As blockages resolve, the system dynamically switches between stages, preventing further threshold violations and effectively managing the queues.
 For shorter blockages ($t \in [437,457]$), Stage 1 alone can support HC traffic at the cost of an increase in accumulated LC packets, yet the blockage is resolved before triggering additional stages.

Next, we evaluate the resilience strategies under different statistical LoS blockage behaviors. We conduct simulations with $N=8$ UEs to analyze the queue evolution across varying mean blockage durations, while maintaining a constant overall LoS blockage probability ($p_b=0.05$ for each link). Through adjustment of the blocking and unblocking rates $\kappa_\mathrm{B}$ and $\mu_\mathrm{B}$, a shorter mean blockage duration indicates frequent blocking and unblocking of LoS paths, while a longer duration implies a lower arrival rate of blockers.
Once again, we compare the activation of different stages of our resilience policy to gain insights into the benefits obtained from each strategy, particularly considering different statistical blockage behaviors. Here, the comparison with the OMA baseline is omitted, as it results in queue instability in this 8-user scenario.

\begin{figure}[t]
    \centering 
    \subfigure[Queuing delay of worst case user]{\label{fig:delay}
    \includegraphics[scale=0.95]{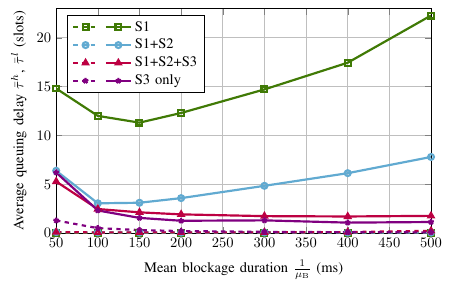}}\vspace{-5pt}
    \subfigure[Proportion of time slots the queue thresholds are exceeded]{\label{fig:overflow}
    \includegraphics[scale=0.95]{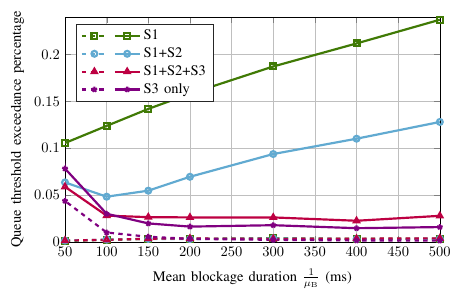}}
    \caption{Average queuing delay and threshold exceedance time of the worst case user as a function of the mean blockage duration $1/{\mu_\mathrm{B}}$, whereby the overall average blockage time of each link remains constant as $p_b=0.05$. We compare different combinations of active resilience stages. Solid lines and dashed lines represent the LC and HC queues, respectively.}
    \label{fig:performance}
\end{figure}

We assess the average worst user queuing delays, along with the fraction of time queue thresholds are exceeded, in Figs. \ref{fig:delay} and \ref{fig:overflow}, respectively.
Results indicate that HC data experiences significantly lower delays and fewer threshold violations compared to LC data. Several factors contribute to the resilience of HC queues: (1) the dual-connectivity of HC data, enabling more reliable decoding in the MC-RSMA scheme, and (2) lower HC queue thresholds that trigger resilience measures early, thereby enhancing responsiveness to imminent congestion.
For LC data, longer, infrequent blockages are more detrimental, leading to higher risks of queue overflow. Conversely, frequent LoS/NLoS transitions enable the system to regularly empty the buffers, thereby preventing excessive packet accumulation. This indicates that adaptive resilience measures become more relevant as failures affect the system over extended periods. Furthermore, for very short blockage durations ($50$ ms) both queuing delay and threshold violations also increase. This is attributed to the delayed blockage detection, which results in more frequent outages caused by undetected blockages during rapid channel fluctuations.

Analyzing Fig. \ref{fig:delay}, when exclusively operating in Stage 1 of our proposed scheme, a substantial rise in queuing delay for LC data is observed when the blockage duration increases beyond 150 ms. In contrast, HC delays remain minimal across all blockage durations due to the multi-connectivity enabled by RSMA.
Incorporating one-sided AP cooperation in Stage 2 results in a considerable delay reduction for LC data, notably, without requiring a reallocation of transmit power. 
While Stage 1 enables blockage robustness only for critical data, the cooperation at Stage 2 expands resilience to LC data. 
Stage 3, in contrast, involves an optimization of transmit power for central decoding under consideration of the detected blockage.
As a result, when considering all three stages within the resilience algorithm, the delay of LC data is further reduced, especially for long blockage durations. Interestingly, the LC queuing delay first decreases with increasing blockage duration until it saturates. This is due to the delayed activation of Stage 3 and the fact that optimized power allocation takes effect with a delay of $n_e$ time steps. This makes full cooperation especially effective for longer-lasting blockages, whereas shorter blockages cause comparatively higher delays.
In comparison, operating permanently in Stage 3 achieves similar performance, with slightly lower delays for blockage durations beyond 100 ms. However, under fast channel fluctuations, Stage 3 alone is less effective, particularly in ensuring low HC delays. Overall, Stage 3 shows a strong performance, yet at the expense of significantly higher computational and coordination complexity, as further analyzed later.

Fig. \ref{fig:overflow} shows a similar behavior with queue threshold violations. 
Despite the HC queue thresholds being notably lower than LC queue thresholds, HC buffers experience very rare threshold violations. Furthermore, incorporating AP cooperation proves to be efficient in preventing buffer overflow where again the benefit is most significant for long blockage durations using all three resilience stages. 
In contrast, operating in Stage 3 alone leads to significantly more HC threshold violations under short blockages (4.38\% with Stage 3 only versus 0.16\% with the three-stage algorithm), underscoring the importance of proactive resilience provided by MC-RSMA.
Both Figs. \ref{fig:delay} and \ref{fig:overflow} reveal that for fast channel LoS/NLoS fluctuations (i.e., short blockage durations), one-sided AP cooperation as in Stage 2 is sufficient, whereas Stage 3, involving a reallocation of resources and switching to central decoding, becomes inevitable for extended blockage durations. Stage 1 remains essential to guarantee HC service continuity regardless of blockage dynamics.

\begin{figure}[t]
    \centering
    \subfigure[Scheme Activation]{\label{fig:activation}
    \includegraphics[scale=0.7]{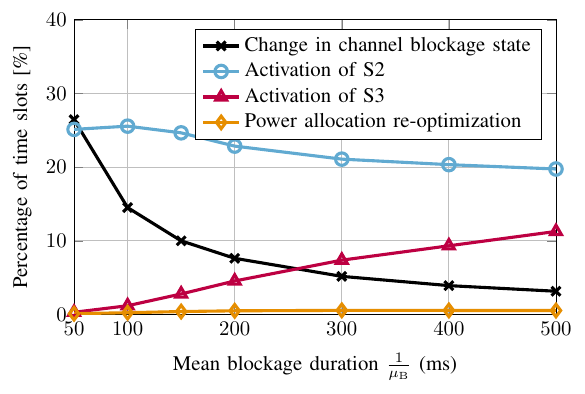}}\vspace{-5pt}
    \subfigure[Operational costs]{\label{fig:costs}
    \hspace{-10pt}
    \includegraphics[scale=0.95]{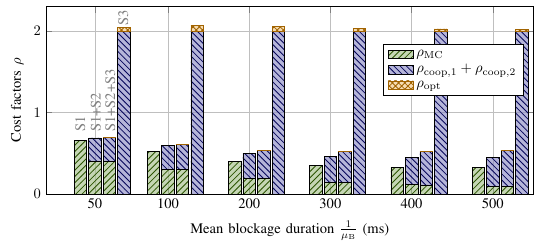}}
    \caption{Proportion of complexity factors and total weighted costs, for different mean blockage duration. From left to right: using S1 only, S1 and S2, all three stages, and S3 only.}
    \label{fig:complexity}
\end{figure}

The benefit from cooperative strategies and transmit power optimization obviously comes at the expense of increased complexity. Therefore, we analyze the event-based activation of different schemes and the entailed operational costs based on failure statistics in Fig. \ref{fig:complexity}. This analysis reveals the importance of incorporating various schemes into the resilience policy and demonstrates the effectiveness of an event-based approach in balancing resilience and operational costs.
 Fig. \ref{fig:activation} provides insights into the activation of cooperation strategies and power allocation adjustments when deploying the full three-stage algorithm. The figure explores these dynamics in relation to the mean blockage duration, or rather the frequency of changes in the blockage state of any UE-AP link.
 In scenarios characterized by rapid LoS/NLoS fluctuations, Stage 3 is rarely activated. This is because short blockages are effectively managed by Stage 1's multi-connectivity and Stage 2's cooperative mechanisms. Since Stage 3 involves more time-consuming resource optimization, it often remains inactive during these short blockage events as blockages resolve before it can be fully utilized.
 As the mean blockage duration increases, the frequency of LoS/NLoS fluctuations decreases, leading to more persistent blockages. In such cases, Stage 3 becomes more active since it has enough time to engage before the blockage is resolved.
 Here, Stage 2 serves primarily as an interim solution until Stage 3's power adjustments take effect.
 Interestingly, the amount of power allocation adjustments remains relatively low across all blockage durations. This is because rapid fluctuations rarely trigger Stage 3, while longer blockages reduce the need for frequent adjustments due to slower channel changes.
 The figure highlights how the event-triggered algorithm dynamically adjusts to different blockage durations. While Stage 1 and Stage 2 effectively handle rapid channel fluctuations, Stage 3 becomes more prominent during longer blockages.
 
 Fig. \ref{fig:costs} illustrates the relative contributions of cost factors $\rho$ associated with multi-connectivity, cooperation, and power optimization across four scenarios: Stage 1 only, Stages 1 and 2 combined, the full three-stage algorithm, and Stage 3 alone. 
 Multi-connectivity is essential for fast LoS/NLoS fluctuations, as its passive robustness effectively handles undetected blockages. When using only Stage 1, the costs driven by multi-connectivity decrease as channel fluctuations slow down, though this comes at the expense of degraded performance (Fig. \ref{fig:performance}).
With Stage 2, cooperation becomes more significant, which introduces additional overhead. During prolonged blockages, cooperation costs dominate, resulting in a relatively stable total cost profile.
Introducing Stage 3 leads to a modest increase in cooperation costs as well as additional power optimization costs. While total costs appear to increase with each stage, they remain manageable relative to the significant delay performance improvements achieved during extended blockages. 
In comparison, operating fully at Stage 3 leads to immense cooperation costs as decoding will take place at the central unit permanently. The optimization costs are also substantially higher when always using Stage 3.
Overall, Fig. \ref{fig:costs} shows that the event-based three-stage algorithm effectively balances costs and performance, activating higher-complexity stages only as needed to ensure stable connectivity across varying blockage scenarios.

\section{Conclusion}\label{sec:conclusion}
In this paper, we studied the efficient design of resilient uplink communication under dynamic LoS blockages. 
Building on a criticality-aware framework that models traffic with mixed-criticality queues, we proposed MC-RSMA as a novel multi-connectivity scheme for robust and efficient uplink transmission. Further leveraging AP cooperation strategies, we developed an event-driven multi-stage resilience framework that combines passive and active robustness with adaptive remediation. From a cross-layer perspective, we formulated two power allocation optimization problems to support fair queue utilization under heterogeneous QoS requirements. 
Simulation results demonstrate the effectiveness of the multi-stage approach and reveal the role of different strategies depending on the statistical behavior of blockages, indicating that:
 \begin{enumerate}
    \item Robust schemes incorporating redundancy and diversity effectively handle frequent, short-term disruptions and uncertainties.
    \item Rare, prolonged failures require adaptive mechanisms, with initial robustness schemes mitigating performance losses before more complex responses take effect.
    \item Criticality-aware, differentiated service treatment is essential to meet user experience and safety demands, while an event-based approach helps balance costs and maintain resource efficiency. 
\end{enumerate}

These findings offer valuable insights for managing various communication disruptions beyond LoS blockages, including traffic overload, intermittent hardware faults, and mobility-induced performance degradation. 
By adjusting resilience stages and triggering policies, the proposed framework can be tailored to diverse network conditions and QoS demands.

The underlying principles are essential for ensuring dependable connectivity in safety-critical applications, where maintaining performance under challenging conditions requires dynamic resource management and differentiated service.

\vspace{-10pt}

\bibliographystyle{./bibliography/IEEEtran}
\bibliography{bibliography/references}

\begin{thebibliography}{10}
\providecommand{\url}[1]{#1}
\csname url@samestyle\endcsname
\providecommand{\newblock}{\relax}
\providecommand{\bibinfo}[2]{#2}
\providecommand{\BIBentrySTDinterwordspacing}{\spaceskip=0pt\relax}
\providecommand{\BIBentryALTinterwordstretchfactor}{4}
\providecommand{\BIBentryALTinterwordspacing}{\spaceskip=\fontdimen2\font plus
\BIBentryALTinterwordstretchfactor\fontdimen3\font minus \fontdimen4\font\relax}
\providecommand{\BIBforeignlanguage}[2]{{%
\expandafter\ifx\csname l@#1\endcsname\relax
\typeout{** WARNING: IEEEtran.bst: No hyphenation pattern has been}%
\typeout{** loaded for the language `#1'. Using the pattern for}%
\typeout{** the default language instead.}%
\else
\language=\csname l@#1\endcsname
\fi
#2}}
\providecommand{\BIBdecl}{\relax}
\BIBdecl

\bibitem{saad20206G}
W.~Saad, M.~Bennis, and M.~Chen, ``{A Vision of 6G Wireless Systems: Applications, Trends, Technologies, and Open Research Problems},'' \emph{IEEE Netw.}, vol.~34, no.~3, pp. 134--142, May/June 2020.

\bibitem{nextGAroadmap}
{Alliance for Telecommunications Industry Solutions}, ``Next {G} alliance report: Roadmap to {6G},'' Tech. Rep., Feb. 2022.

\bibitem{IMT-2030}
\emph{IMT-2030 Framework: Recommendation ITU-R M.2160-0}, International Telecommunication Union (ITU), ITU-R Working Party 5D, Nov. 2023.

\bibitem{madni2009resilience}
A.~M. Madni and S.~Jackson, ``{Towards a Conceptual Framework for Resilience Engineering},'' \emph{IEEE Syst. J.}, vol.~3, no.~2, pp. 181--191, 2009.

\bibitem{maccartney2017blockage}
G.~R. MacCartney, T.~S. Rappaport, and S.~Rangan, ``{Rapid Fading Due to Human Blockage in Pedestrian Crowds at 5G Millimeter-Wave Frequencies},'' in \emph{IEEE Global Commun. Conf. (GLOBECOM)}, Singapore, Dec. 2017, pp. 1--7.

\bibitem{khaloopour2024resilience}
L.~Khaloopour, Y.~Su, F.~Raskob, T.~Meuser, R.~Bless, L.~Janzen, K.~Abedi, M.~Andjelkovic, H.~Chaari, P.~Chakraborty, M.~Kreutzer, M.~Hollick, T.~Strufe, N.~Franchi, and V.~Jamali, ``{Resilience-by-Design in 6G Networks: Literature Review and Novel Enabling Concepts},'' \emph{IEEE Access}, vol.~12, pp. 155\,666--155\,695, 2024.

\bibitem{mahmood2024resilient}
N.~H. Mahmood, S.~Samarakoon, P.~Porambage, M.~Bennis, and M.~Latva-aho, ``{Resilient-By-Design: A Resiliency Framework for Future Wireless Networks},'' \emph{arXiv preprint arXiv:2410.23203}, 2024.

\bibitem{perez2023multiAntenna}
D.~E. Perez, O.~L.~A. Lopez, and H.~Alves, ``{Robust Downlink Multi-Antenna Beamforming With Heterogenous CSI: Enabling eMBB and URLLC Coexistence},'' \emph{IEEE Trans. Wireless Commun.}, vol.~22, no.~6, pp. 4146--4157, June 2023.

\bibitem{aboagye2024multiband}
S.~Aboagye, M.~Amin~Saeidi, H.~Tabassum, Y.~Tayyar, E.~Hossain, H.-C. Yang, and M.-S. Alouini, ``{Multi-Band Wireless Communication Networks: Fundamentals, Challenges, and Resource Allocation},'' \emph{IEEE Trans. Commun.}, vol.~72, no.~7, pp. 4333--4383, July 2024.

\bibitem{lin2023Rel19}
X.~Lin, ``{The Bridge Toward 6G: 5G-Advanced Evolution in 3GPP Release 19},'' \emph{arXiv preprint arXiv:2312.15174}, 2023.

\bibitem{khoury2014multi}
M.~Khoury and S.~Bullock, ``{Multi-level resilience: reconciling robustness, recovery and adaptability from a network science perspective},'' \emph{Int. J. Adapt. Resilient Auton. Syst. (IJARAS)}, vol.~5, no.~4, pp. 34--45, 2014.

\bibitem{punzo2020resilience}
G.~Punzo, A.~Tewari, E.~Butans, M.~Vasile, A.~Purvis, M.~Mayfield, and L.~Varga, ``{Engineering Resilient Complex Systems: The Necessary Shift Toward Complexity Science},'' \emph{IEEE Syst. J.}, vol.~14, no.~3, pp. 3865--3874, Sept. 2020.

\bibitem{sterbenz2010resilience}
J.~P. Sterbenz, D.~Hutchison, E.~K. {\c{C}}etinkaya, A.~Jabbar, J.~P. Rohrer, M.~Sch{\"o}ller, and P.~Smith, ``{Resilience and survivability in communication networks: Strategies, principles, and survey of disciplines},'' \emph{Comput. Netw.}, vol.~54, no.~8, pp. 1245--1265, June 2010.

\bibitem{kaada2022resilience}
S.~Kaada, M.~L. Alberi~Morel, G.~Rubino, and S.~Jelassi, ``{Resilience analysis and quantification method for 5G-Radio Access Networks},'' in \emph{13th Int. Conf. Netw. Future (NoF)}, Ghent, Belgium, Oct. 2022, pp. 1--9.

\bibitem{reifert2022comeback}
R.-J. Reifert, S.~Roth, A.~A. Ahmad, and A.~Sezgin, ``{Comeback Kid: Resilience for Mixed-Critical Wireless Network Resource Management},'' \emph{IEEE Trans. Veh. Technol.}, vol.~72, no.~12, pp. 16\,177--16\,194, 2023.

\bibitem{shui2024design}
T.~Shui and W.~Saad, ``{Design and Analysis of Resilient Vehicular Platoon Systems over Wireless Networks},'' in \emph{IEEE Global Commun. Conf. (GLOBECOM)}, 2024.

\bibitem{li2023trafficresilience}
R.~Li, B.~Decocq, A.~Barros, Y.-P. Fang, and Z.~Zeng, ``{Estimating 5G Network Service Resilience Against Short Timescale Traffic Variation},'' \emph{IEEE Trans. Netw. Service Manag.}, vol.~20, no.~3, pp. 2230--2243, 2023.

\bibitem{gerasimenko2019multiconn}
M.~Gerasimenko, D.~Moltchanov, M.~Gapeyenko, S.~Andreev, and Y.~Koucheryavy, ``{Capacity of Multiconnectivity mmWave Systems With Dynamic Blockage and Directional Antennas},'' \emph{IEEE Trans. Veh. Technol.}, vol.~68, no.~4, pp. 3534--3549, Apr. 2019.

\bibitem{rahman2019beamswitch}
A.~U. Rahman and G.~Ghatak, ``{A Beam-Switching Scheme for Resilient mm-Wave Communications With Dynamic Link Blockages},'' in \emph{Int. Symp. Modeling Opt. Mob., Ad Hoc, Wireless Netw. (WiOPT)}, Avignon, France, June 2019, pp. 1--6.

\bibitem{kumar2021CoMP}
D.~Kumar, J.~Kaleva, and A.~T\"olli, ``{Blockage-Aware Reliable mmWave Access via Coordinated Multi-Point Connectivity},'' \emph{IEEE Trans. Wireless Commun.}, vol.~20, no.~7, pp. 4238--4252, July 2021.

\bibitem{reifert2023CoNOMA}
R.-J. Reifert, H.~Dahrouj, and A.~Sezgin, ``{Extended Reality via Cooperative NOMA in Hybrid Cloud/Mobile-Edge Computing Networks},'' \emph{IEEE Internet Things J.}, pp. 1--1, 2023.

\bibitem{barbarossa2019resilient}
N.~di~Pietro, M.~Merluzzi, E.~C. Strinati, and S.~Barbarossa, ``{Resilient design of 5G mobile-edge computing over intermittent mmWave links},'' \emph{arXiv preprint arXiv:1901.01894}, 2019.

\bibitem{karacora2024intermittency}
Y.~Karacora, A.~Umra, and A.~Sezgin, ``{Intermittency Versus Path Loss in RIS-aided THz Communication: A Data Significance Approach},'' in \emph{IEEE Int. Conf. Commun. (ICC)}, 2024, pp. 3414--3419.

\bibitem{karacora2024THzRIS}
------, ``{Robust Communication Design in RIS-Assisted THz Channels},'' \emph{arXiv preprint arXiv:2411.10524}, 2024.

\bibitem{bassoli2021why6G}
R.~Bassoli, F.~H. Fitzek, and E.~C. Strinati, ``{Why do we need 6G?}'' \emph{ITU J. Future Evolving Technol.}, vol.~2, no.~6, pp. 1--31, Sept. 2021.

\bibitem{maham2020CoMPuplink}
B.~Maham and P.~Popovski, ``{Capacity Analysis of Coordinated Multipoint Reception for mmWave Uplink With Blockages},'' \emph{IEEE Trans. Veh. Technol.}, vol.~69, no.~12, pp. 16\,299--16\,303, Dec. 2020.

\bibitem{abbasi2022transmission}
O.~Abbasi and H.~Yanikomeroglu, ``{Transmission scheme, detection and power allocation for uplink user cooperation with NOMA and RSMA},'' \emph{IEEE Trans. Wireless Commun.}, vol.~22, no.~1, pp. 471--485, Jan. 2023.

\bibitem{reifert2022MC_RSMA}
R.-J. Reifert, S.~Roth, A.~A. Ahmad, and A.~Sezgin, ``{Energy Efficiency in Rate-Splitting Multiple Access with Mixed Criticality},'' in \emph{{IEEE Int. Conf. Commun. Workshops (ICC Workshops)}}, Seoul, Rep. of Korea, May 2022, pp. 681--686.

\bibitem{reifert2023AoI}
R.-J. Reifert, S.~Roth, and A.~Sezgin, ``{Optimizing the Age of Information in Mixed-Critical Wireless Communication Networks},'' in \emph{IEEE Int. Conf. Commun. (ICC)}, Rome, Italy, June 2023, pp. 1682--1687.

\bibitem{park2021nonlinearAoI}
T.~Park, W.~Saad, and B.~Zhou, ``{On the Minimization of Non-Linear Age of Information in the Internet of Things},'' in \emph{IEEE Int. Conf. Commun.}, Montreal, QC, Canada, June 2021, pp. 1--6.

\bibitem{popovski2018H_NOMA}
P.~Popovski, K.~F. Trillingsgaard, O.~Simeone, and G.~Durisi, ``{5G Wireless Network Slicing for eMBB, URLLC, and mMTC: A Communication-Theoretic View},'' \emph{IEEE Access}, vol.~6, pp. 55\,765--55\,779, Sept. 2018.

\bibitem{mao2022rate}
Y.~Mao, O.~Dizdar, B.~Clerckx, R.~Schober, P.~Popovski, and H.~V. Poor, ``{Rate-splitting multiple access: Fundamentals, survey, and future research trends},'' \emph{IEEE Commun. Surveys Tuts.}, 2022.

\bibitem{yang2019uplinkRSMA}
Z.~Yang, M.~Chen, W.~Saad, W.~Xu, and M.~Shikh-Bahaei, ``{Sum-Rate Maximization of Uplink Rate Splitting Multiple Access (RSMA) Communication},'' in \emph{IEEE Global Commun. Conf. (GLOBECOM)}, Waikoloa, HI, USA, Dec. 2019, pp. 1--6.

\bibitem{ahmad2019BSbreakdown}
A.~A. Ahmad, J.~Kakar, R.-J. Reifert, and A.~Sezgin, ``{UAV-Assisted C-RAN with Rate Splitting Under Base Station Breakdown Scenarios},'' in \emph{IEEE Int. Conf. Commun. Workshops (ICC Workshops)}, Shanghai, China, May 2019, pp. 1--6.

\bibitem{karacora2022ratesplitting}
Y.~Karacora and A.~Sezgin, ``{Rate-Splitting Enabled Multi-Connectivity in Mixed-Criticality Systems},'' in \emph{IEEE Int. Conf. Commun. (ICC)}, Rome, Italy, June 2023, pp. 5340--5345.

\bibitem{jain2019block}
I.~K. Jain, R.~Kumar, and S.~S. Panwar, ``{The impact of mobile blockers on millimeter wave cellular systems},'' \emph{IEEE J. Sel. Areas Commun.}, vol.~37, no.~4, pp. 854--868, Apr. 2019.

\bibitem{kumar2016M2M}
A.~Kumar, A.~Abdelhadi, and C.~Clancy, ``{A delay-optimal packet scheduler for M2M uplink},'' in \emph{IEEE Military Commun. Conf. (MILCOM)}, 2016, pp. 295--300.

\bibitem{chaccour2020risk}
C.~Chaccour, M.~N. Soorki, W.~Saad, M.~Bennis, and P.~Popovski, ``{Risk-Based Optimization of Virtual Reality over Terahertz Reconfigurable Intelligent Surfaces},'' in \emph{IEEE Int. Conf. Commun. (ICC)}, Dublin, Ireland, June 2020, pp. 1--6.

\bibitem{neely2010introduction}
M.~J. Neely, ``{Introduction to Queues},'' in \emph{{Stochastic Network Optimization with Application to Communication and Queueing Systems}}.\hskip 1em plus 0.5em minus 0.4em\relax Springer, 2010, pp. 15--28.

\bibitem{littlesLaw}
J.~D.~C. Little, ``{A Proof for the Queuing Formula: {$L= \lambda W$}},'' \emph{Operations Research}, vol.~9, no.~3, pp. 383--387, 1961.

\bibitem{fettweis2021tactile}
G.~P. Fettweis and H.~Boche, ``{6G: The Personal Tactile Internet -- And Open Questions for Information Theory},'' \emph{IEEE BITS Inf. Theory Mag.}, vol.~1, no.~1, pp. 71--82, Sept. 2021.

\bibitem{shen2018fractional}
K.~Shen and W.~Yu, ``{Fractional programming for communication systems--Part I: Power control and beamforming},'' \emph{IEEE Trans. Signal Process.}, vol.~66, no.~10, pp. 2616--2630, May 2018.

\bibitem{cvx}
M.~Grant and S.~Boyd, ``{CVX}: Matlab software for disciplined convex programming, version 2.1,'' \url{http://cvxr.com/cvx}, Mar. 2014.

\bibitem{gray1993quantizer}
R.~Gray and T.~Stockham, ``{Dithered quantizers},'' \emph{IEEE Trans. Inf. Theory}, vol.~39, no.~3, pp. 805--812, May 1993.

\bibitem{cover1999elements}
T.~M. Cover and J.~A. Thomas, \emph{{Elements of Information Theory}}.\hskip 1em plus 0.5em minus 0.4em\relax John Wiley \& Sons, 1991.

\bibitem{fritzsche2013backhaul}
R.~Fritzsche, E.~Ohlmer, and G.~P. Fettweis, ``{Where to Predict the Channel in Cooperative Cellular Networks with Backhaul Delays?}'' in \emph{9th Int. ITG Conf. Sys., Commun. Cod. (SCC)}, Munich, Germany, Jan. 2013, pp. 1--6.

\end{thebibliography}

\end{document}